\documentclass[epj,nopacs]{svjour}

\usepackage{amsmath, amssymb}
\usepackage[german,english]{babel}
\usepackage[utf8]{inputenc}
\usepackage{graphicx}
\usepackage{mathtools}
\usepackage{nicefrac}
\usepackage{color}
\usepackage[numbers,sort&compress]{natbib}

\newcommand*{\coord}[1]{\mathbf{#1}}
\newcommand*{\vecr}{\coord{r}}

\newcommand*{\crea}[1]{\hat{#1}^{\dagger}}
\newcommand*{\anni}[1]{\hat{#1}^{\vphantom{\dagger}}}

\newcommand*{\binteg}[3]{\int^{\mathrlap{#3}}_{\mathrlap{#2}}\ud{#1}\,}
\newcommand*{\integ}[1]{\!\int\!\ud{#1}\:}
\newcommand*{\iinteg}[2]{\integ{#1}\!\!\integ{#2}}

\DeclareMathOperator{\Ei}{Ei}
\DeclareMathOperator{\Order}{\mathcal{O}}

\DeclareMathOperator{\trace}{tr}
\DeclareMathOperator{\Trace}{Tr}

\DeclarePairedDelimiter{\abs}{\lvert}{\rvert}
\DeclarePairedDelimiter{\av}{\langle}{\rangle}

\DeclarePairedDelimiterX\braket[2]{\langle}{\rangle}{#1\delimsize\vert#2}
\DeclarePairedDelimiterX\brakket[3]{\langle}{\rangle}{#1\delimsize\vert#2\delimsize\vert#3}
\DeclarePairedDelimiterX\ket[1]{\lvert}{\rangle}{#1}
\DeclarePairedDelimiterX\bra[1]{\langle}{\rvert}{#1}

\newcommand*{\du}{\partial}
\newcommand*{\eqspace}{\phantom{{} = {}}}
\newcommand*{\e}{\mathrm{e}}
\newcommand*{\half}{\frac{1}{2}}
\newcommand*{\im}{\textrm{i}}
\newcommand*{\isDefinedAs}{\coloneqq}
\newcommand*{\nhalf}{\nicefrac{1}{2}}

\newcommand*{\thalf}{\tfrac{1}{2}}
\newcommand*{\ud}{\mathrm{d}}

\newcounter{gapSave}
\newcounter{potSave}
\newcounter{eqSave}

\usepackage{tikz}
\usetikzlibrary{positioning, calc}
\usetikzlibrary{decorations, decorations.pathmorphing, decorations.markings}
\usetikzlibrary{arrows, shapes.geometric}
\usetikzlibrary{patterns}

\tikzset{/pgf/decoration/.cd,
    angle step/.initial=10,
}
\newdimen\tmpdimen
\pgfdeclaredecoration{complete sines}{initial}
{
    \state{initial}[
        width=+0pt,
        next state=move,
        persistent precomputation={
            \pgfmathparse{\pgfkeysvalueof{/pgf/decoration/angle step}}%
            \let\anglestep=\pgfmathresult%
            \let\currentangle=\pgfmathresult%
            \pgfmathsetlengthmacro{\pointsperanglestep}%
                {(\pgfdecoratedremainingdistance/%
                  (round(\pgfdecoratedinputsegmentlength/\pgfdecorationsegmentlength)+0.5)/360*\anglestep}%
        }] {}
    \state{move}[width=+\pointsperanglestep, next state=draw]{
        \pgfpathmoveto{\pgfpointorigin}
    }
    \state{draw}[width=+\pointsperanglestep, switch if less than=1.25*\pointsperanglestep to final, 
        persistent postcomputation={
        \pgfmathparse{mod(\currentangle+\anglestep, 360)}%
        \let\currentangle=\pgfmathresult%
    }]{%
        \pgfmathsin{+\currentangle}%
        \tmpdimen=\pgfdecorationsegmentamplitude%
        \tmpdimen=\pgfmathresult\tmpdimen%
        \divide\tmpdimen by2\relax%
        \pgfpathlineto{\pgfqpoint{0pt}{\tmpdimen}}%
    }
    \state{final}{
        \ifdim\pgfdecoratedremainingdistance>0pt\relax
            \pgfpathlineto{\pgfpointdecoratedpathlast}
        \fi
   }
}

\tikzset{
  bare photon/.style={
    decoration={complete sines, amplitude=0.15cm, segment length=0.25cm},
    decorate
  },
  dressed photon/.style={
    bare photon, double distance=0.5pt
  },
  bare fermion/.style={
    decoration={
      markings, mark= at position 0.5 with {
        \node[transform shape, xshift=-0.5mm, fill=black, inner sep=1pt, draw, isosceles triangle]{};
      }
    },
    postaction=decorate
  },
  dressed fermion/.style={
    bare fermion, double, double distance=1pt
  },
  vertex/.style={
    circle, inner sep =1.25pt, minimum size=3pt, fill=black, outer sep=0
  },
  HugVertex/.style={
    circle, inner sep =1.75pt, minimum size=6pt, fill=black, outer sep=0
  }
}

\allowdisplaybreaks[4]

\usepackage{hyperref}

\begin{document}

\title{Approximate energy functionals for one-body reduced density matrix functional theory from many-body perturbation theory.}
\titlerunning{Approximate energy functionals for 1RDM functional theory MBPT.}

\author{Klaas J. H. Giesbertz\inst{1} \and Anna--Maija Uimonen\inst{2} \and Robert van Leeuwen\inst{2}}
\authorrunning{K.J.H. Giesbertz \and A. Uimonen \and R. van Leeuwen}

\institute{Theoretical Chemistry, Faculty of Exact Sciences, VU University, De Boelelaan 1083, 1081 HV Amsterdam, The Netherlands, \email{klaas.giesbertz@gmail.com} \and Department of Physics, Nanoscience Center, University of Jyväskylä, P.O. Box 35, 40014 Jyväskylä, Survontie 9, Jyväskylä, Finland, \email{robert.vanleeuwen@jyu.fi}}

\mail{K.J.H. Giesbertz}

\date{\today}

\abstract{
We develop a systematic approach to construct energy functionals of the one-particle reduced density matrix (1RDM)
for equilibrium systems at finite temperature. The starting point of our formulation is the grand potential $\Omega [\vec{G}]$ regarded
as variational functional of the Green's function $G$ based on diagrammatic many-body perturbation theory and for which we consider
either the Klein or Luttinger--Ward form. By restricting the input Green's function to be one-to-one related to a set on one-particle
reduced density matrices (1RDM)  this functional becomes a functional of the 1RDM. 
To establish the one-to-one mapping we use
that, at any finite temperature and for a given 1RDM $\vec{\gamma}$ in a finite basis, there exists a non-interacting system with a spatially non-local potential $v[\vec{\gamma}]$ 
which reproduces the given 1RDM. The corresponding set of non-interacting Green's functions defines the variational domain of the functional $\Omega$.
In the zero temperature limit we obtain an energy functional $E[\vec{\gamma}]$ which by minimisation yields an approximate  ground state 1RDM and energy.  
As an application of the formalism we use the Klein and Luttinger--Ward functionals in the GW-approximation compute 
the binding curve of a model hydrogen molecule using an extended Hubbard Hamiltonian. We compare further to the case in which we evaluate
the functionals on a Hartree--Fock and a Kohn--Sham Green's function. We find that the Luttinger--Ward version of the functionals performs the best and is able to
reproduce energies close to the GW energy which corresponds to the stationary point.
}

\maketitle

\section{Introduction}
\label{sec:intro}

One-body reduced density matrix (1RDM) functional theory provides an interesting alternative to density function theory (DFT), which holds the promise to alleviate many of the current practical failures of DFT. Especially the inherent better capability of 1RDM functional theory to deal with strongly correlated systems, e.g.\ the breaking of chemical bonds~\cite{GritsenkoPernalBaerends2005,RohrPernalGritsenko2008,PirisLopezRuiperez2011,RuiperezPirisUgalde2013}, Mott insulator transitions~\cite{SharmaDewhurstLathiotakis2008,SharmaDewhurstShallcross2013} and the fractional quantum Hall effect~\cite{ToloHarju2010}, are an encouragement to invest even further in the development of 1RDM functionals.

Currently, two main strategies are followed in the development of new 1RDM functionals. The first strategy uses the fact that the exact 1RDM functional for two-electron systems is available which is used as a paradigm to device general $N$-electron functionals~\cite{PhD-Buijse1991,BuijseBaerends2002,GritsenkoPernalBaerends2005,RohrPernalGritsenko2008,MentelMeerGritsenko2014}. The other main approach is to reconstruct the two-body reduced density matrix (2RDM) from the 1RDM guided by $N$-representability conditions~\cite{Mazziotti2001,PirisOtto2003,Piris2006,PirisMatxainLopez2009,PirisLopezRuiperez2011,PirisMatxainLopez2013,Piris2014,Piris2017}. A more extensive overview of different approaches can be found in Ref.~\cite{PernalGiesbertz2015}.

A disadvantage of the current approaches is that they do not provide a systematic route towards improved 1RDM functionals. In many-body perturbation theory (MBPT), however, such a route is available by including an increasing amount of terms in the perturbation expansion~\cite{FetterWalecka1971,StefanucciLeeuwen2013}. These terms in the perturbation expansion are conveniently represented by Feynman diagrams. It would therefore advantages to connect the MBPT framework to 1RDM functional theory, since this would allow one to use the MBPT functionals for the construction of increasingly more accurate 1RDM functionals in a systematic manner. This approach has already been heavily pursued in DFT with some considerable success~\cite{Furche2001b,DahlenBarth2004,DahlenLeeuwenBarth2006,Furche2008,HellgrenBarth2010,HeselmannGorling2010,EshuisBatesFurche2012,BatesFurche2013,BleizifferHeselmannGorling2013,HellgrenCarusoRohr2015}, so the additional flexibility of 1RDM functional theory to a further improvement of the results, especially when strong correlations play an important role. Due to the peculiar properties of 1RDM functional theory, however, the connection to MBPT is not so trivial as for DFT. In DFT one can use the Kohn--Sham (KS) system~\cite{KohnSham1965b}, from which a non-interacting Green's function can be extracted to be inserted into the MBPT functional. The construction of a similar KS system in 1RDM functional theory --- a non-interacting system with the same 1RDM as the interacting system --- leads to a completely degenerate orbital spectrum, since the orbitals are fractionally occupied~\cite{Gilbert1975,Pernal2005,GiesbertzBaerends2010}. For Coulomb systems we know at least that infinitely many orbitals are fractionally occupied~\cite{Friesecke2003} and there is strong evidence that they all are~\cite{GiesbertzLeeuwen2013a,GiesbertzLeeuwen2013b,GiesbertzLeeuwen2014}. The corresponding Green's function of the non-interacting system in 1RDM functional theory has therefore all its poles located at the same position. Such an unphysical Green's function is bound to lead to nonsensical results when inserted directly into an MBPT functional.

The degeneracy of the orbital energies in 1RDM functional theory seems to pose a serious problem if one wants to use MBPT for the construction of functionals. However, the massive degeneracy of the orbital energies is caused by the fact that we evaluate the system at zero temperature and can be lifted by elevating the temperature to any finite value, $T > 0$. The degeneracy problem can therefore circumvented by performing the calculation at a finite temperature. Within a finite basis set, 1RDM functional theory allows for a completely rigorous foundation in which all required functionals are nicely differentiable and no $v$-representability issues arise~\cite{GiesbertzRuggenthaler2017}.
The zero temperature limit, $T \to 0$, can be taken at the end of the calculation. An additional advantage is that also the finite-temperature MBPT formalism is more reliable, since finite-temperature MBPT does not rely on adiabatically turning on the interaction, which leads to problems for zero-temperature MBPT in the case of level crossings~\cite{FetterWalecka1971,StefanucciLeeuwen2013}. We will elaborate later in this article more on this extraction process of the non-interacting Green's function from the 1RDM.

The extracted Green's function can now be deployed in the MBPT framework which can be done in several ways. The most straightforward option is to use the extracted Green's function as a zeroth order Green's function in the MBPT formalism. Unfortunately, this leads to complicated expressions: the effective potential needed to force the non-interacting system to have the same 1RDM as the interacting system needs to be subtracted from all occurrences of the self-energy~\cite{GorlingLevy1993,GorlingLevy1994,PhD-Baldsiefen2012,BaldsiefenGross2012a,BlochlPruschkePotthoff2013}. This means that also the energy expression depends on the effective potential itself, so becomes a severe complication. More importantly, following this approach one only obtains a reformulation of MBPT in terms of a different (less convenient) zeroth order Green's function. The results will therefore be the same as for an MBPT calculation with the same diagrams, so the only achievement would be a more complicated formalism, if one aims for fully self-consistent solutions.

A more viable approach is to use variational MBPT functionals $\Omega[\vec{G}]$ for the total energy~\cite{AlmbladhBarthLeeuwen1999,AryasetiawanMiyakeTerakura2002,MiyakeAryasetiawanKotani2002,DahlenBarth2004,DahlenLeeuwenBarth2006}. The total energy from these functionals are relatively stable with respect to variations in the Green's function, due their variational property $\delta\Omega/\delta\vec{G} = 0$ when $\vec{G}$ is obtained self-consistently form the Dyson equation. This has the advantage that one does not need to solve the full Dyson equation, but that it is sufficient to use a reasonable approximate Green's function to obtain a reasonably accurate value for the total energy. Such an approximate Green's function can be the KS Green's function or the non-interacting Green's function extracted from the 1RDM.

We have different variational functionals at our disposal and popular forms are the Luttinger--Ward~\cite{LuttingerWard1960} and Klein~\cite{Klein1961} functionals. Especially the Klein functional is quite popular, as it yields the simplest expression for the energy~\cite{DahlenBarth2004,DahlenLeeuwenBarth2006}. The Klein functional can be reformulated in terms of the exchange-correlation kernel to yield the adiabatic connection fluctuation-dissipation expression~\cite{NiquetFuchsGonze2003,HeselmannGorling2010,EshuisBatesFurche2012,CarusoRohrHellgren2013}. Though the Luttinger--Ward functional results in a more complicated expression for the energy than the Klein functional, it has the advantage of superior variational properties. We will take both functionals in consideration in this article, but limit ourselves to one of the simplest perturbative expansions: the GW. The performance of both the Klein and Luttinger--Ward functionals in the GW approximation will be tested on an extensions of the two-site Hubbard model which is capable to give a good representation of the ground state of the hydrogen molecule. To achieve this, the extended two-site Hubbard model also takes the intersite interactions into account and the interaction between the electrons and the nuclei.

The paper is divided as follows. In Sec.~\ref{sec:theory} we present the general theory of the variational functionals that we use. In Sec.~\ref{sec:Gs} we show how to construct a non-interacting Green's function $\vec{G}_s[\vec{\gamma}]$ that yields a prescribed 1RDM $\vec{\gamma}$. In Sec.~\ref{sec:model} we present the details for the molecular model that we use to test our functionals. In Sec.~\ref{sec:inputG} we describe the various input Green's functions that we use as input for the variational functionals. In Sec.~\ref{sec:variational} we perform variational calculations on the molecular model and finally, in Sec.~\ref{sec:conclusion} we present our conclusions.

\section{Variational functionals}
\label{sec:theory}
The required theoretical framework for the use of variational MBPT functionals has basically already been presented in the framework of DFT in Ref.~\cite{DahlenBarth2004}. The only difference is that in the framework, we now use non-local potentials to keep the complete 1RDM fixed at all interaction strengths. The procedure is very analogous to the derivation of the Luttinger--Ward functional~\cite[Sec.~11.4]{StefanucciLeeuwen2013}. For completeness, we will provide here a short overview. Consider the \begin{equation}
\hat{H}^{\lambda} = \sum_{ij} \bigl(k_{ij} + v_{ij}^{\lambda}\bigr)\crea{c}_i\anni{c}_j +
\lambda\sum_{ijkl}w_{ijkl}\crea{c}_i\crea{c}_j\anni{c}_k\anni{c}_l ,
\end{equation}
where $k_{ij}$ and $w_{ijkl}$ are are one-and two-body matrix elements. The term $\vec{v}^{\lambda}$ is a potential to keep the 1RDM identical at all interaction strengths $\lambda$ to the fully interacting 1RDM, $\vec{\gamma}^{\lambda} = \vec{\gamma}^1 = \vec{\gamma}$. The potential at $\lambda = 1$ is therefore the real external potential $\vec{v}^1 = \vec{v}_{\text{ext}}$. In the framework of KS-DFT the potential $\vec{v}^{\lambda}$ would be local, i.e.\ diagonal in the coordinate representation or site basis, and only the density would be kept fixed at all interaction strengths, $\vec{n}^{\lambda} = \vec{n}$. At $\lambda = 0$, the Hamiltonian becomes non-interacting $\hat{H}_s \isDefinedAs \hat{H}^0$ and the its corresponding Green's function $\vec{G}_s$ has a particularly simple form (see Sec.~\ref{sec:Gs}). The corresponding potential to keep the 1RDM or density of the non-interacting system identical to the one of the non-interacting system is denote by $v_s \isDefinedAs v^0$.

At general $\lambda$, the Green's function is related to the non-interacting Green's function via the Dyson equation
\begin{equation}\label{eq:Dyson}
\vec{G}^{\lambda} = \vec{G}_s + \vec{G}_s\vec{\tilde{\Sigma}}^{\lambda}\vec{G}^{\lambda} ,
\end{equation}
where $\vec{\tilde{\Sigma}}^{\lambda} \isDefinedAs \vec{\Sigma}^{\lambda} + \vec{v}^{\lambda} - \vec{v}_s$.
To relate the grand potentials of the interacting and non-interacting system, we will use the fundamental theorem of calculus. So we first work out out the derivative of the grand potential with respect to the interaction strength
\begin{equation}
\frac{\ud\Omega^{\lambda}}{\ud\lambda}
= \av[\bigg]{\frac{\ud\hat{H}^{\lambda}}{\ud\lambda}}
= \frac{1}{\beta}\Trace\biggl\{\biggl(\frac{\ud \vec{v}^{\lambda}}{\ud\lambda} + \frac{\vec{\Sigma}^{\lambda}}{2\lambda}\biggr)\vec{G}^{\lambda}\biggr\} ,
\end{equation}
where $\Trace$ indicates a summation over all indices as well as a summation over the Matsubara frequencies.
We now use the $\Phi$-functional which can be constructed by summing over all irreducible self-energy diagrams closed with an additional Green's function~\cite{LuttingerWard1960}. Each of these closed diagrams is multiplied by the pre-factor $1/(2n)$, where $n$ denotes the number of interaction lines
\begin{equation}
\Phi^{\lambda}[\vec{G}]
= \sum_{n,k}\frac{1}{2n}\Trace\bigl\{\vec{\Sigma}^{\lambda(n)}_k\vec{G}\bigr\} .
\end{equation}
The self-energy derived from the $\Phi$ functional as
\begin{equation}\label{eq:SigmaDef}
\vec{\Sigma}[\vec{G}] \isDefinedAs \frac{\delta\Phi}{\delta\vec{G}},
\end{equation}
leads to a conserving approximation for $\vec{\Sigma}$~\cite{BaymKadanoff1961,Baym1962}. Its derivative with respect to the coupling strength is readily worked out as
\begin{equation}
\frac{\ud\Phi^{\lambda}}{\ud\lambda}
= \frac{1}{2\lambda}\Trace\bigl\{\vec{\Sigma}^{\lambda}\vec{G}^{\lambda}\bigr\} +
\Trace\biggl\{\vec{\Sigma}^{\lambda}\frac{\ud\vec{G}^{\lambda}}{\ud\lambda}\biggr\} ,
\end{equation}
which allows us to rewrite the derivative of the grand potential as
\begin{equation}
\beta\frac{\ud\Omega^{\lambda}}{\ud\lambda}
= \frac{\ud}{\ud\lambda}\biggl(\Phi^{\lambda} - \Trace\bigl\{\vec{\Sigma}^{\lambda}\vec{G}^{\lambda}+ \ln\bigl(\vec{1} - \vec{G}_s\vec{\tilde{\Sigma}}^{\lambda}\bigr)\bigr\}\biggr) .
\end{equation}
The last term is readily verified by working out the derivative of $\Trace\bigl\{\ln\bigl(\vec{1} - \vec{G}_s\vec{\tilde{\Sigma}}^{\lambda}\bigr)\bigr\}$~\cite{StefanucciLeeuwen2013}.
Integrating from the non-interacting system to the fully interacting one, we find that their grand potentials are related as
\begin{align}\label{eq:LWfunctional}
\Omega_{\text{LW}}[\vec{G}] ={}& \Omega_s
+ \trace\bigl\{(\vec{v}_{\text{ext}} - \vec{v}_s)\vec{\gamma}\bigl\} \\*
&{} + \frac{1}{\beta}\Bigl(\Phi[\vec{G}] - \Trace\bigl\{\vec{\tilde{\Sigma}}\vec{G} +
\ln\bigl(\vec{1} - \vec{G}_s\vec{\tilde{\Sigma}}\bigr)\bigr\}\Bigr) , \notag
\end{align}
where
\begin{align}\label{eq:modSigmaDef}
\vec{\tilde{\Sigma}} \isDefinedAs \vec{\tilde{\Sigma}}^1
= \vec{\Sigma} + \vec{v}_{\text{ext}} - \vec{v}_s
\end{align}
and $\trace$ only sum over matrix indices. By regarding the grand potential as a functional of the one-body Green's function we have retrieved the Luttinger--Ward (LW) functional $\Omega_{\text{LW}}[\vec{G}]$, for an arbitrary non-interacting reference Green's function $\vec{G}_s$. Note that this expression assumes that we are keeping the complete 1RDM constant for varying interaction strength. If we were only to keep the density fixed, $\vec{\gamma}$ should be replaced by $\vec{n}$ and the potential $\vec{v}_{\text{ext}}$ and $\vec{v}_s$ would be local. Note that apart from the allowance of an arbitrary spatially non-local potential instead of the true external potential to be able to work with more general $\vec{G}_s$, the result 
in~\eqref{eq:LWfunctional} is
identical to that obtained originally by Luttinger and Ward~\cite{LuttingerWard1960}.

When integrating over the interaction strength, we have assumed that the chemical potential is also kept fixed at its interacting value. This condition fixes the constant in $\vec{v}^{\lambda}$, i.e. $\trace\bigl\{\vec{v}^{\lambda}\bigr\}$, as it should be chosen such that the particle number remains constant at varying interaction strengths. This is readily guaranteed by assuring that the Luttinger--Ward functional yields correct number of particles, $N = -\du\Omega_{\text{LW}}/\du\mu$. This is particularly convenient as the term $\mu N$ can now easily be eliminated to obtain the total energy.

The Luttinger--Ward functional is variational in the sense that when $\vec{G}$ solves the Dyson equation~\eqref{eq:Dyson} with $\vec{\Sigma} = \delta\Phi/\delta\vec{G}$, perturbations in $\Omega_{\text{LW}}$ vanish to first order
\begin{equation}
\label{LWvar}
\frac{\delta\Omega_{\text{LW}}}{\delta\vec{G}} = 0 .
\end{equation}
This does not guarantee that the Luttinger--Ward functional achieves its minimum at the solution of the Dyson equation, but only that it is stationary.

One can construct infinitely many different expressions for the grand potential with a different functional dependence of the Green's function which both yield the correct value of the grand potential and are stationary at the solution of the Dyson equation~\cite{StefanucciLeeuwen2013}. A popular alternative variational functional is the one due to Klein~\cite{Klein1961}
\begin{multline}\label{eq:KleinFunctional}
\Omega_{\text{K}}[\vec{G}] \isDefinedAs
\Omega_s + \trace\bigl\{(\vec{v}_{\text{ext}} - \vec{v}_s)\vec{\gamma}\bigr\} \\*
{}+ \frac{1}{\beta}\Bigl[\Phi[\vec{G}] + \Trace\bigl\{\ln(\vec{G}\vec{G}_s^{-1}) + \vec{1} - \vec{G}\vec{G}_s^{-1}\bigr\}\Bigr] .
\end{multline}
The Klein functional is simpler to evaluate, as there is no explicit reference to the self-energy. The price we pay for this simplification is that the Klein functional is less stable to perturbations than the Luttinger--Ward functional. Though the first order variations vanish at the solution of the Dyson equation, the higher order terms typically have larger amplitudes. This higher sensitivity to the input Green's function of the Klein functional has been demonstrated for atoms~\cite{DahlenBarth2004} and diatomic molecules~\cite{DahlenLeeuwenBarth2006}. The simplicity of the Klein functional is especially apparent if we insert the non-interacting Green's function $\vec{G}_s$ into the Klein functional. All the terms within $\Trace\{\cdot\}$ disappear and only the $\Phi$-functional remains as a non-trivial part.

Finally, we address some practical applications of the variational functionals. First of all let us consider that the stationary point of the functional
is attained for a Green's function $\vec{G}$ but that we insert a different input Green's function $\tilde{\vec{G}}$. Due to the stationarity property (\ref{LWvar})
(and similarly for the Klein functional) we have, with $\Delta \vec{G} = \tilde{\vec{G}} - \vec{G}$ that
\begin{equation}
\Omega [\tilde{\vec{G}}] = \Omega [\vec{G}] + \half \int \!\!\frac{\delta^2 \Omega}{\delta \vec{G} \delta \vec{G}} [\vec{G}] \Delta \vec{G} \Delta \vec{G} + \Order \bigl( (\Delta \vec{G})^3\bigr) ,
\end{equation}
where the second order derivative in the integrand on the right hand side is a tensor contracted with terms $\Delta \vec{G}$ and we further imply a double integration of imaginary times.
We therefore see that if we make an error $\Delta \vec{G}$ in the Green's function then the error in $\Omega$ is, due to its variational property, only quadratic in this error.
This is a very useful property as it ensure that we may obtain good values of the grand potential from rather simple input Green's functions, provided they are close enough
in some sense to the stationary point. The size of the error also depends on the value of second derivative in the integral, which is different for the LW and Klein forms
of the energy functional and in practice the variational property of the LW form is found to be superior~\cite{DahlenBarth2004, DahlenLeeuwenBarth2006}. 
Another way to use the energy functionals is by looking for the stationary point on a restricted domain. For example by inserting Green's functions from a general non-interacting system
with a local potential we obtain a density functional~\cite{DahlenBarth2004,DahlenLeeuwenBarth2006} since the potential is in one-to-one correspondence
with a density by means of the Hohenberg--Kohn (or Mermin) theorem. In this work we extend this to non-local potentials since at finite temperature and in a finite basis it can be proven
that any ensemble N-representable 1RDM is $v$-representable by a non-local potential~\cite{GiesbertzRuggenthaler2017}. Using this property, we can consider the variational domain of $\Omega [\vec{G}]$
to be the set of Green's functions of non-interacting systems at a finite temperature with a non-local potential. By varying over the non-local potentials we vary
over a domain of 1RDMs and we are effectively using a 1RDM functional the stationary point of which yields an approximate 1RDM. This approach will be
used in the present work. Note that by restricting the variational domain we will not recover the exact Green's function anymore at the stationary point of the variational equations. This
would even be the case had we used the exact $\Phi$-functional. We further remark that a different way to construct density matrix functionals on the basis of Green's functions is given in Ref.~\cite{BlochlPruschkePotthoff2013} and the proposed approximate scheme in Section IV~C of Ref.~\cite{BlochlPruschkePotthoff2013} coincides with our use of the Luttinger--Ward functional. The method that we present here is close in spirit to the one presented in Ref.~\cite{BarthDahlenLeeuwen2005}.

\section{Construction of $\vec{G}_s [\vec{\gamma}]$}
\label{sec:Gs}
In this Section we outline how we can construct a non-interacting system at finite temperature that produces a given 1RDM. The
Green's function $\vec{G}_s [\vec{\gamma}]$ of this system produces functionals of the 1RDM when inserted in the variational Luttinger--Ward and Klein functionals. These 1RDM functional will be 
studied in greater detail in later sections.

The Matsubara Green's function of a non-interacting system consisting of bosonic\slash{}fermionic particles can be shown to be of the form~\cite{FetterWalecka1971,StefanucciLeeuwen2013}
\begin{equation}\label{eq:Gs}
G_{s,kl}(\tau) = -\im\,\delta_{kl}\bigl[\theta(\tau)\bar{f}_k \pm \theta(-\tau)f_k\bigr]\e^{-\tau\epsilon^{\text{M}}_k},
\end{equation}
where $f_k \isDefinedAs (\e^{\beta\epsilon^{\text{M}}_k} \mp 1)^{-1}$ is the Bose\slash{}Fermi distribution function, $\bar{f}_k \isDefinedAs 1 \pm f_k$ and $\beta \isDefinedAs 1/T$ is the inverse temperature. The Heaviside step function is defined as
\begin{equation}
\theta(\tau) = \begin{cases}
1	&\text{if $\tau > 0$} \\
0	&\text{if $\tau < 0$} .
\end{cases}
\end{equation}
The Matsubara energies are obtained by subtracting the chemical potential from the eigenvalues of the one-body Hamiltonian, $\epsilon^{\text{M}}_k \isDefinedAs \epsilon_k - \mu$, where the chemical potential can be adjusted such that the system has the desired number of particles
\begin{equation}
N = \sum_kf_k = \sum_k\frac{1}{\e^{\beta(\epsilon_k - \mu)} \mp 1} .
\end{equation}
The corresponding non-interacting 1RDM is readily extracted from the Green's function as
\begin{equation}
\gamma_{s,kl} = \pm\im G_{s,kl}(0^-) = f_k\,\delta_{kl} .
\end{equation}
Since the 1RDM is diagonal in the eigenbasis of the one-body Hamiltonian, the Bose\slash{}Fermi distribution functions are the natural occupation numbers.

Now let us limit the discussion to electrons, i.e.\ fermions. It is clear from the form of the Fermi function that it satisfies the strict inequalities $0 < f_k < 1$ at finite temperatures, so the occupation numbers are purely fractional at $T > 0$. In the limit $T \to 0$, however, the Fermi function collapses to a Heaviside function and the occupation numbers become integers 0 or 1 depending on the sign of $\epsilon^{\text{M}}_k$. For $\epsilon^{\text{M}}_k = 0$ the value of the occupation number is undetermined at $T = 0$ and can be anything between zero and one, $0 \leq f_k \leq 1$, in principle. However, by considering how the $\epsilon^{\text{M}}_k \to 0$ in the zero temperature limit, the occupation numbers will have a well defined value in the $T \to 0$ limit even if $\epsilon^{\text{M}}_k \to 0$.

\begin{figure}
  \includegraphics[width=\columnwidth]{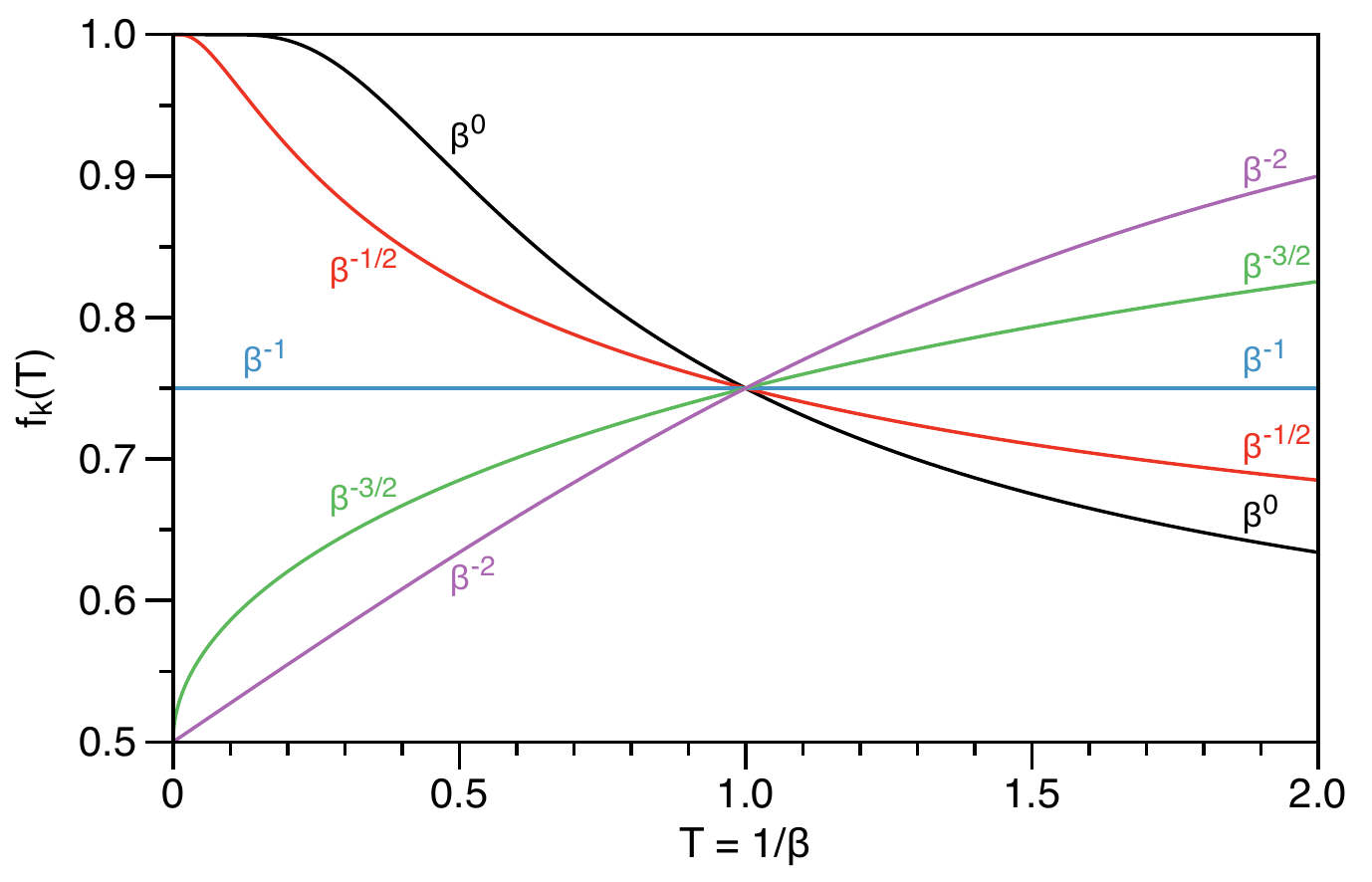}
  \caption{Behaviour of the occupation numbers calculated for different temperature dependence of the Matsubara energy, $\epsilon^{\text{M}}_k = \beta^{-n}\ln(3)$. The $\beta^0$ curve (black) corresponds to the more accustomed temperature independent behaviour.}
  \label{fig:occConverge}
  \includegraphics[width=\columnwidth]{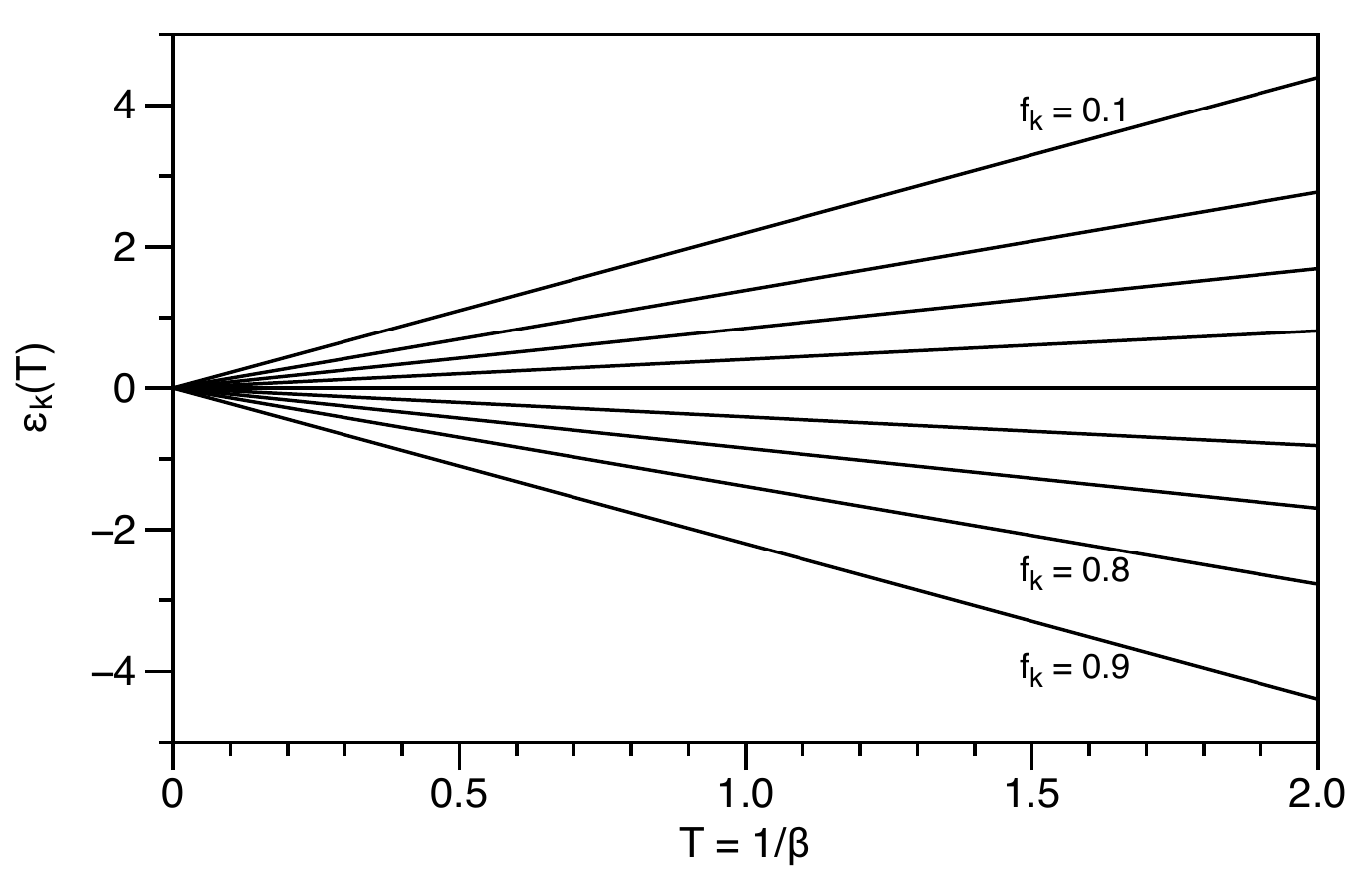}
  \caption{Behaviour of the Matsubara energies as a function of the temperature for fractional occupation numbers. These occupation numbers are $0.1$ apart from each other.}
  \label{fig:epsConverge}
\end{figure}

Suppose we have a given 1RDM and want to find the Matsubara Hamiltonian, $\hat{H}^{\text{M}} \isDefinedAs \hat{H} - \mu\hat{N}$, which yields this 1RDM. Since the eigenstates of the one-body Hamiltonian and the 1RDM coincide, it is clear that the eigenfunctions of the one-body Hamiltonian should be the natural orbitals (eigenfunctions of the 1RDM). The corresponding eigenvalues are readily obtained by inverting the Fermi distribution
\begin{equation}
\epsilon^{\text{M}}_k = \frac{1}{\beta}\ln\biggl(\frac{1 - f_k}{f_k}\biggr) = \frac{1}{\beta}\ln\biggl(\frac{\bar{f}_k}{f_k}\biggr) .
\end{equation}
From this expression it is clear that if the Matsubara energies in the $\beta \to \infty$ limit decay as $\epsilon^{\text{M}}_k \sim 1/\beta$ that the occupation number will have a definite value $0 < f_k < 1$. If its decay is slower it will end up at one of the integer values depending on its sign. If the Matsubara energy decays faster than $1/\beta$, the corresponding occupation number converges to $f_k = \nhalf$. The behaviour of the occupation number in the $T \to 0$ for these various decay rates are depicted in Fig.~\ref{fig:occConverge}. The collapse of the Matsubara energy spectrum for fractional occupation numbers in the $T \to 0$ limit is demonstrated in Fig.~\ref{fig:epsConverge}.

\section{Two-orbital model for H$_2$}
\label{sec:model}
To keep the calculations as simpel as possible, we limit ourselves to a two-orbital model for the hydrogen molecule. 
Let us start from the full interacting Hamiltonian for H$_2$
\begin{equation}
\hat{H} = \sum_{\crampedclap{ij,\sigma}}h_{ij}\crea{c}_{i\sigma}\anni{c}_{j\sigma} +
\thalf\sum_{\crampedclap{ijkl,\sigma\sigma'}}w_{ijkl}\crea{c}_{i\sigma}\crea{c}_{j\sigma'}\anni{c}_{k\sigma'}\anni{c}_{l\sigma} ,
\end{equation}
where the one-electron matrix elements,
\begin{equation}
h_{ij} \isDefinedAs \brakket{\varphi_i}{\hat{h}}{\varphi_j} = \integ{\vecr}\varphi^*_i(\vecr)\hat{h}\varphi^{\vphantom{*}}_j(\vecr) ,
\end{equation}
contain the kinetic energy and external potential. The two-electron matrix elements, $\vec{w}$, describe the Coulomb interaction between the electrons and are given as
\begin{equation}
w_{ijkl}
\isDefinedAs \iinteg{\vecr}{\vecr'}\frac{\varphi^*_i(\vecr)\varphi^*_j(\vecr')
\varphi^{\vphantom{*}}_k(\vecr')\varphi^{\vphantom{*}}_l(\vecr)}{\abs{\vecr - \vecr'}}
= [il \vert jk] .
\end{equation}
In this equation we also used the quantum chemical notation~\cite{SzaboOstlund1989} which regards the two-electron integral as a weighted overlap between charge distribution $\varphi^*_i(\vecr)\varphi^{\vphantom{*}}_l(\vecr)$ and $\varphi^*_j(\vecr')\varphi^{\vphantom{*}}_k(\vecr')$. When we work with orthonormal orbitals, we can make the tight-binding approximation to the two-electron integrals
\begin{align}
[il \vert kl] \approx \delta_{il}\delta_{kl}[ii \vert jj] .
\end{align}
The underlying idea is that the charge distributions $\varphi^*_i(\vecr)\varphi^{\vphantom{*}}_l(\vecr)$ integrates to zero for $i \neq l$ and therefore leads to a much smaller value of the two-electron integral than the diagonal term $i = l$. If we further only use one orbital per hydrogen atom, the Hamiltonian simplifies in the tight-binding approximation to
\begin{align}
\hat{H} ={}& \alpha\hat{N} +
t\sum_{\sigma}\bigl(\crea{c}_{1\sigma}\anni{c}_{2\sigma} + \crea{c}_{2\sigma}\anni{c}_{1\sigma}\bigr)  +
w\,\hat{n}_1\hat{n}_2 \notag \\
{}&+U\bigl(\hat{n}_{1\uparrow}\hat{n}_{1\downarrow} + \hat{n}_{2\uparrow}\hat{n}_{2\downarrow}\bigr) +
\frac{\bar{U}}{2}\sum_{i,\sigma}\crea{c}_{i\sigma}\crea{c}_{i\sigma}\anni{c}_{i\sigma}\anni{c}_{i\sigma} ,
\end{align}
where $\hat{n}_{i\sigma} \isDefinedAs \crea{c}_{i\sigma}\anni{c}_{i\sigma}$, $\hat{n}_i \isDefinedAs \hat{n}_{i\uparrow} + \hat{n}_{i\downarrow}$ and $\hat{N} = \hat{n}_1 + \hat{n}_2$.
This Hamiltonian can be regarded as an extension of the Hubbard model for two sites. The normal Hubbard Hamiltonian only includes the hopping term between the two sites, whose strength is governed by $t$, and the on-site interaction, whose strength is set by $U$. We additionally include a term depending on the number of particles with strength $\alpha$. This term does not have an influence the eigenstates, since they are eigenfunctions of the total number of particles, but does change the corresponding eigenvalues, however, so is important to recover the correct electronic energy. Apart from the on-site interaction, we also include the interaction between the electrons if they reside on different sites and its strength is determined by the parameter $w$.

The last term in the extended Hubbard Hamiltonian is a self-interaction term and does not contribute in principle.
However, when we construct energy functionals from the perturbation expansion, the expansion for the $\Phi$ functional cannot not always directly be associated to a proper expansion of an anti-symmetric 2-body Green’s function.
In that case the energy functional is not guaranteed to be self-interaction free and $\bar{U}$ typically does make a contribution. The GW (RPA) approximation considered in this article is an example of such an expansion which is not self-interaction free, hence the final result depends on $\bar{U}$.
In this work, we will consider two different values for $\bar{U}$. The value $\bar{U} = U$ corresponds to a spin-independent interaction, such as the non-relativistic Coulomb interaction used in molecular Hamiltonians. The other sensible value to use is $\bar{U} = 0$, which explicitly eliminates the self-interaction at the level of the Hamiltonian. This is the usual choice in the Hubbard model.

To determine the values of the extended Hubbard parameters, we choose normalised 1s orbitals located at each hydrogen atom to construct our basis
\begin{align}
\chi_1(\vecr) &= N_{\zeta}\e^{-\zeta\abs{\vecr - \coord{R}_A}} ,	&
\chi_2(\vecr) &= N_{\zeta}\e^{-\zeta\abs{\vecr - \coord{R}_B}} ,
\end{align}
where the normalization factor is given by $N_{\zeta} = \sqrt{\zeta^3/\pi}$. The exponent $\zeta$ will be variationally optimized to obtain the lowest energy. In the dissociation limit $R \isDefinedAs \abs{\coord{R}_A - \coord{R}_B} \to \infty$ we have that $\zeta = 1$. A localised orthonormal basis is readily construct by Löwdin orthogonalization~\cite{Lowdin1950}, $\vec{\varphi} = \vec{\chi}\vec{S}^{-\nhalf}$, where $S_{ij} = \braket{\chi_i}{\chi_j}$ is the overlap matrix of the non-orthogonal basis. Using this Löwdin orthogonalised basis, the one-electron matrix elements in the extended Hubbard Hamiltonian can be determined as
\begin{subequations}
\label{eq:LowdinOneElectron}
\begin{align}
\alpha &= \brakket{\varphi_1}{\hat{h}}{\varphi_1}
= \frac{\brakket{\chi_1}{\hat{h}}{\chi_1} - s\brakket{\chi_1}{\hat{h}}{\chi_2}}{1 - s^2} , \\
t &= \brakket{\varphi_1}{\hat{h}}{\varphi_2}
= \frac{\brakket{\chi_1}{\hat{h}}{\chi_2} - s\brakket{\chi_1}{\hat{h}}{\chi_1}}{1 - s^2} ,
\end{align}
\end{subequations}
where $s \isDefinedAs \braket{\chi_1}{\chi_2}$ denotes the overlap. For more details one these expressions and explicit forms in terms of the orbital exponent $\zeta$ and the internuclear distance $R$, consult Appendix~\ref{ap:matElems}.

The two-electron matrix elements can be expressed as
\begin{subequations}
\label{eq:LowdinTwoElectron}
\begin{align}
U &= (11 \vert 11) + \frac{s^2}{2(1 - s^2)}\bigl[(11 \vert 11) - (11 \vert 22)\bigr] , \\
w &= (11 \vert 22) + \frac{s^2}{2(1 - s^2)}\bigl[(11 \vert 22) - (11 \vert 11)\bigr] ,
\end{align}
\end{subequations}
where we need the two-electron matrix elements in the non-orthogonal 1s-basis
\begin{equation}
(ij \vert kl) \isDefinedAs \iinteg{\vecr}{\vecr'}
\frac{\chi^*_i(\vecr)\chi^{\vphantom{*}}_j(\vecr)\,\chi^*_k(\vecr')\chi^{\vphantom{*}}_l(\vecr')}{\abs{\vecr - \vecr'}} .
\end{equation}
These two-electron matrix elements for the 1s-basis are worked out explicitly in terms of $\zeta$ and $R$ in Appendix~\ref{ap:matElems}.

\begin{figure}[t]
  \includegraphics[width=\columnwidth]{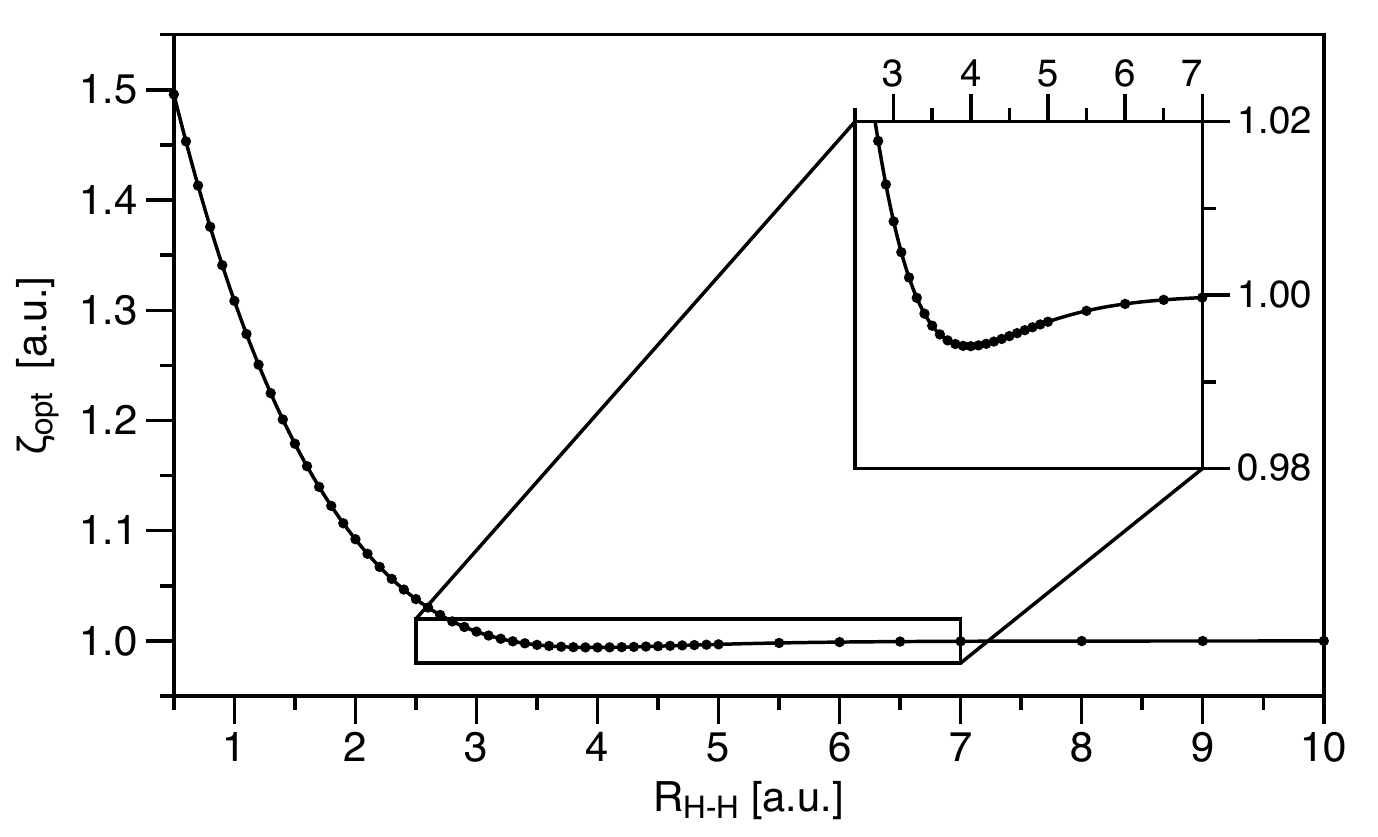}
  \caption{The exponent of the orbital, $\zeta$, as a function of the internuclear distance, $R_{\text{H--H}}$.}
  \label{fig:orbExpo}
\end{figure}

Due to the limited dimension of the Hilbert space, the extended two-site Hubbard model is easy to solve exactly. In the two-particle sector we obtain the following triplet solutions
\begin{align}
&\crea{c}_{1\downarrow}\crea{c}_{2\downarrow}\ket{} , &
&\tfrac{1}{\sqrt{2}}\bigl(\crea{c}_{1\uparrow}\crea{c}_{2\downarrow} + 
\crea{c}_{1\downarrow}\crea{c}_{2\uparrow}\bigr)\ket[\big]{} , &
&\crea{c}_{1\uparrow}\crea{c}_{2\uparrow}\ket{} ,
\end{align}
with the eigenvalue $E = 2\alpha + w$. The singlet states can be subdivided according to their parity. There is only one ungerade singlet state, $\tfrac{1}{\sqrt{2}}\bigl(\crea{c}_{1\uparrow}\crea{c}_{1\downarrow} - \crea{c}_{2\uparrow}\crea{c}_{2\downarrow}\bigr)\ket[\big]{}$, with the eigenvalue $E = 2\alpha + U$. There are two gerade singlet states which can be expressed as
\begin{multline}
\tfrac{1}{\sqrt{2}}\bigl[
\cos(\alpha_{\pm})\bigl(\crea{c}_{1\uparrow}\crea{c}_{2\downarrow} - \crea{c}_{1\downarrow}\crea{c}_{2\uparrow}\bigr) \\
{} - \sin(\alpha_{\pm})\bigl(\crea{c}_{1\uparrow}\crea{c}_{1\downarrow} + \crea{c}_{2\uparrow}\crea{c}_{2\downarrow}\bigr)
\bigr]\ket[\big]{} ,
\end{multline}
where the angles $\alpha_{\pm}$ can be determined from the equation
\begin{equation}
\tan(\alpha_{\pm}) = \frac{w - U}{4t} \pm \sqrt{\cramped{\biggl(\frac{w - U}{4t}\biggr)^{\mathrlap{2}}} + 1} .
\end{equation}
The corresponding electronic energies of the gerade singlet states are
\begin{equation}
E_{\pm} = 2\alpha + \frac{w + U}{2} \pm \sqrt{\cramped{\biggl(\frac{w - U}{2}\biggr)^{\mathrlap{2}}} + 4t^2} .
\end{equation}
The energy $E_-$ is the ground state energy which is minimized by optimizing the value for the orbital exponent $\zeta$. The optimal value, $\zeta_{\text{opt}}$, is shown in Fig.~\ref{fig:orbExpo}. The value of the exponent goes to 1 when stretching the bond. This is expected, since in the dissociation limit the system consists of two separated hydrogen atoms. The asymptotic value is approached from below, since the 1s orbital needs to become more diffuse to facilitate binding. Binding would be more efficiently achieved by allowing the 1s orbital to polarize towards the other hydrogen atom, for example by mixing in the $2p_z$ orbital. By allowing for polarization, the reduction in the exponent would be less.

\begin{figure}[t]
  \includegraphics[width=\columnwidth]{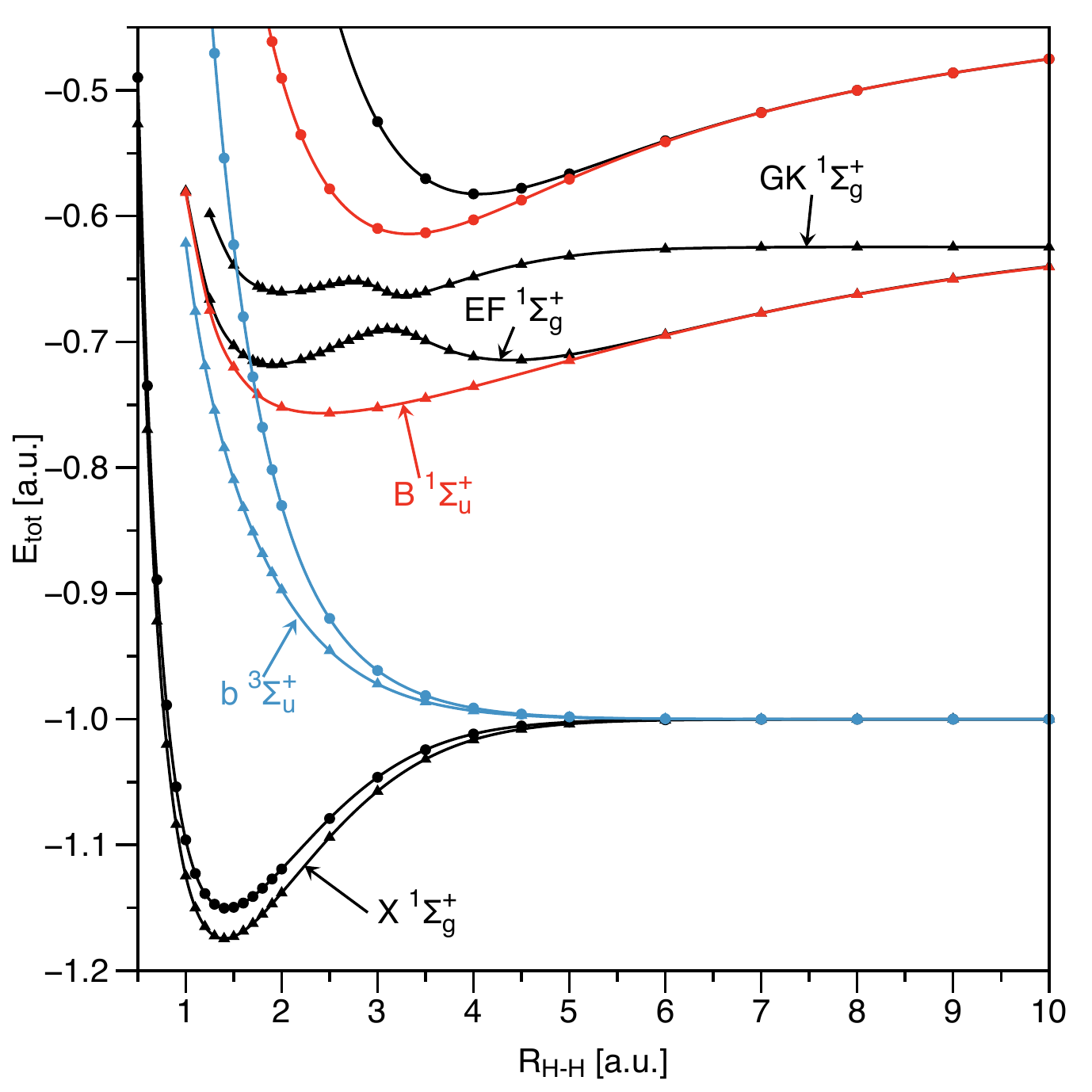}
  \caption{Comparison of the extended Hubbard energies (circles) with the exact potential energy curves (triangles)~\cite{KolosWolniewicz1965,KolosWolniewicz1966,KolosWolniewicz1969,WolniewiczDressler1985}. The energy of the (first) two\slash{}three $^1\Sigma_g^+$ states are shown in black, the energy of the (lowest) $^3\Sigma_u^+$ state in blue and the (lowest) $^1\Sigma_u^+$ state in red.}
\end{figure}

The values for the total energy from our extended two-site Hubbard model are compared to the accurate values obtained for the non-relativistic hydrogen molecule in the Born--Oppenheimer approximation by Kołos and Wolniewicz~\cite{KolosWolniewicz1965,KolosWolniewicz1966,KolosWolniewicz1969,WolniewiczDressler1985}. Since the free parameter $\zeta$ is optimized for the ground state, $X\,^1\Sigma_g^+$, it is reproduced quite accurately. The energy of the triplet $b\,^3\Sigma_u^+$ state is of less quality around the equilibrium bond length $R_e = 1.4$ Bohr, but behaves very well upon dissociation. The other excited states are upshifted and there minima are located at too large bond lengths. Nevertheless the overall shape of the $B\,^1\Sigma_u^+$ state is reasonable and correctly becomes degenerate with the $EF\,^1\Sigma_g^+$ state in the dissociation limit. The double-well structure of the $EF\,^1\Sigma_g^+$ is completely absent, which is no surprise, since the higher lying $GK\,^1\Sigma_g^+$ required for the necessary avoided crossing is not present in our simple two-orbital model.

\section{Input Green's functions}
\label{sec:inputG}
In the following sections we will assess the performance of the Klein and the Luttinger--Ward (LW) functionals using the Green's functions as input: the Kohn--Sham (KS), the Hartree--Fock (HF) and 1RDM Green's function. These non-interacting Green's functions are based on a one-body Hamiltonian. Since we only use two basis functions in the model, the symmetry of the system dictates that the eigenfunctions of the one-body Hamiltonian need to be
\begin{equation}
\phi_{g/u}(\vecr) = \frac{\varphi_1(\vecr) \pm \varphi_2(\vecr)}{\sqrt{2}}
= \frac{\chi_1(\vecr) \pm \chi_2(\vecr)}{\sqrt{2(1 \pm s)}}.
\end{equation}
So the Matsubara non-interacting Green's function~\eqref{eq:Gs} are also diagonal in this symmetry adapted basis. As we restrict the optimisation to symmetry adapted solutions, we do not allow for broken symmetry states to handle strong correlation effects. The full burden is placed on the functional.

We will limit the discussion to neutral H$_2$, so $N = 2(f_g + f_u) = 2$, which immediately implies that the chemical potential should be chosen such that $\epsilon^{\text{M}}_g + \epsilon^{\text{M}}_u = 0$, so the chemical potential in the non-interacting system is required to be
\begin{equation}
\mu_s = \frac{\epsilon_g + \epsilon_u}{2} .
\end{equation}

\subsection{Hartree--Fock approximation}
One of the simplest perturbation expansions involves only the Hartree--Fock (HF) diagrams
\begin{align}
-\Phi^{\text{HF}} &= \half
\vcenter{\hbox{%
\begin{tikzpicture}[node distance=1]
  \node [vertex] (a) {};
  \node [vertex, right=0.75 of a] (b) {};
  \draw [bare photon] (a) -- (b);
  \draw [dressed fermion] (b) arc (180:-180:0.4);
  \draw [dressed fermion] (a) arc (360:0:0.4);
  \node [vertex] at (a) {};
  \node [vertex] at (b) {};
\end{tikzpicture}%
}}
+ \half
\vcenter{\hbox{%
\begin{tikzpicture}[node distance=1]
  \node [vertex] (a) {};
  \node [vertex, right=of a.center, anchor=center] (b) {};
  \draw [bare photon] (a) -- (b);
  \draw [dressed fermion] (b.center) arc (0:180:0.5);
  \draw [dressed fermion] (a.center) arc (180:360:0.5);
  \node [vertex] at (a) {};
  \node [vertex] at (b) {};
\end{tikzpicture}%
}} \notag \\
&= -\frac{\beta}{2}\Bigl[\bigl(U + 2w\bigr)\tilde{n}^2 - w \tilde{f}^2\Bigr],
\end{align}
where $\tilde{n} \isDefinedAs f_g + f_u = N/2 = 1$ and $\tilde{f} \isDefinedAs f_g - f_u$ denotes the difference in occupation of the gerade and ungerade orbital. The corresponding HF self-energy is particularly simple, since it is local in time
\begin{align}\label{eq:SigmaHF}
\Sigma^{\text{HF}}_{kl}(z_1,z_2)	&=
\vcenter{\hbox{%
\begin{tikzpicture}[node distance=1]
  \node [vertex] (a) {};
  \node [vertex, above=0.5 of a] (b) {};
  \draw [bare photon] (a) -- (b);
  \draw [dressed fermion] (b) arc (-90:270:0.35);
  \node [vertex] at (b) {};
  \node [anchor=mid, left = 0.15 of a.center] {$z_1\,k$};
  \node [anchor=mid, right = 0.15 of a.center] {$l\,z_2$};
\end{tikzpicture}%
}} +
\vcenter{\hbox{%
\begin{tikzpicture}[node distance=1]
  \node [vertex] (a) {};
  \node [vertex, right=of a] (b) {};
  \draw [dressed fermion] (b) -- (a);
  \draw [bare photon] (a) to [out=90, in=90, looseness=1.5] (b);
  \node [anchor=mid, left = 0.15 of a.center] {$z_1\,k$};
  \node [anchor=mid, right = 0.15 of b.center] {$l\,z_2$};
\end{tikzpicture}%
}} \notag \\
&= v^{\text{HF}}_{kl}\delta(z_1,z_2) ,
\end{align}
where the HF potential for our simple model system is diagonal in the symmetry-adapted basis and the non vanishing elements are (see supplement)
\begin{subequations}\label{eq:HFpotential}
\begin{align}
v^{\text{HF}}_{gg} = v^{\text{HF}}_{\bar{g}\bar{g}}
&= \thalf\bigl(U + 2w\bigr)\tilde{n} - \thalf w\tilde{f} , \\
v^{\text{HF}}_{uu} = v^{\text{HF}}_{\bar{u}\bar{u}}
&= \thalf\bigl(U + 2w\bigr)\tilde{n} + \thalf w\tilde{f} .
\end{align}
\end{subequations}
The lowest energy is obtained by fully occupying the $\sigma_g$ orbitals, $\tilde{f} = 1$. As the HF potential differs by $w$ between the gerade and ungrade orbitals, so the HF gap is increased by $w$ with respect to the gap of the non-interacting system. Hence, we have
\setcounter{gapSave}{\value{equation}}
\begin{subequations}
\begin{equation}\label{eq:HFgap}
\epsilon^{\text{HF}} = \epsilon_u^{\text{HF}} - \epsilon_g^{\text{HF}} = -2t + w .
\end{equation}
\end{subequations}
As the HF potential is completely fixed by~\eqref{eq:HFpotential}, the chemical potential in the HF approximation becomes
\begin{equation}\label{eq:HFchemPot}
\mu_{\text{HF}}\bigl(\tilde{n} = 1\bigr)
= \half\trace\bigl\{\vec{h} + \vec{v}^{\text{HF}}\bigr\}
= \alpha + \frac{U + 2w}{2} .
\end{equation}
Later, we will demonstrate that the correlation part of the self-energy does not introduce any additional contributions to the constant in the potential, so $\mu = \mu_{\text{HF}}$. Hence, the effective potential in~\eqref{eq:LWfunctional} becomes
\setcounter{potSave}{\value{equation}}
\begin{subequations}\label{eq:pots}
\begin{equation}
\vec{v}_s^{\text{HF}} = \vec{v}_{\text{ext}} + \vec{v}^{\text{HF}} .
\end{equation}
\end{subequations}

\subsection{The KS Green's function}
\label{sec:KSgreen}
The KS system has a particularly simple realisation in this two-orbital model. As the KS potential is local, its can only have diagonal elements in the site basis. As the KS should not break the symmetry of the system, it needs to be equal on both sites, so it can only be a constant shift. As the only constant contributions come from the external potential and the Hartree part, we need to set
\setcounter{eqSave}{\value{equation}}
\setcounter{equation}{\value{potSave}}
\begin{subequations}
\stepcounter{equation}
\begin{equation}
\vec{v}^{\text{KS}}_s = \mu\vec{1} = \vec{v}_{\text{ext}} + \frac{U + 2w}{2}\vec{1}
\end{equation}
\end{subequations}
in order ot have $\mu_{\text{KS}} = \mu$.
Since $t < 0$ for any finite $R_{\text{H--H}}$, occupying the $\sigma_g$ orbital will always lead to the lowest KS energy. The gap in the KS system is now completely fixed by hopping matrix elements
\setcounter{equation}{\value{gapSave}}
\begin{subequations}
\stepcounter{equation}
\begin{equation}
\epsilon^{\text{KS}} = \epsilon_u^{\text{KS}} - \epsilon_g^{\text{KS}} = -2t .
\end{equation}
\end{subequations}

\subsection{The 1RDM Green's function}
\begin{figure}[t]
  \includegraphics[width=\columnwidth]{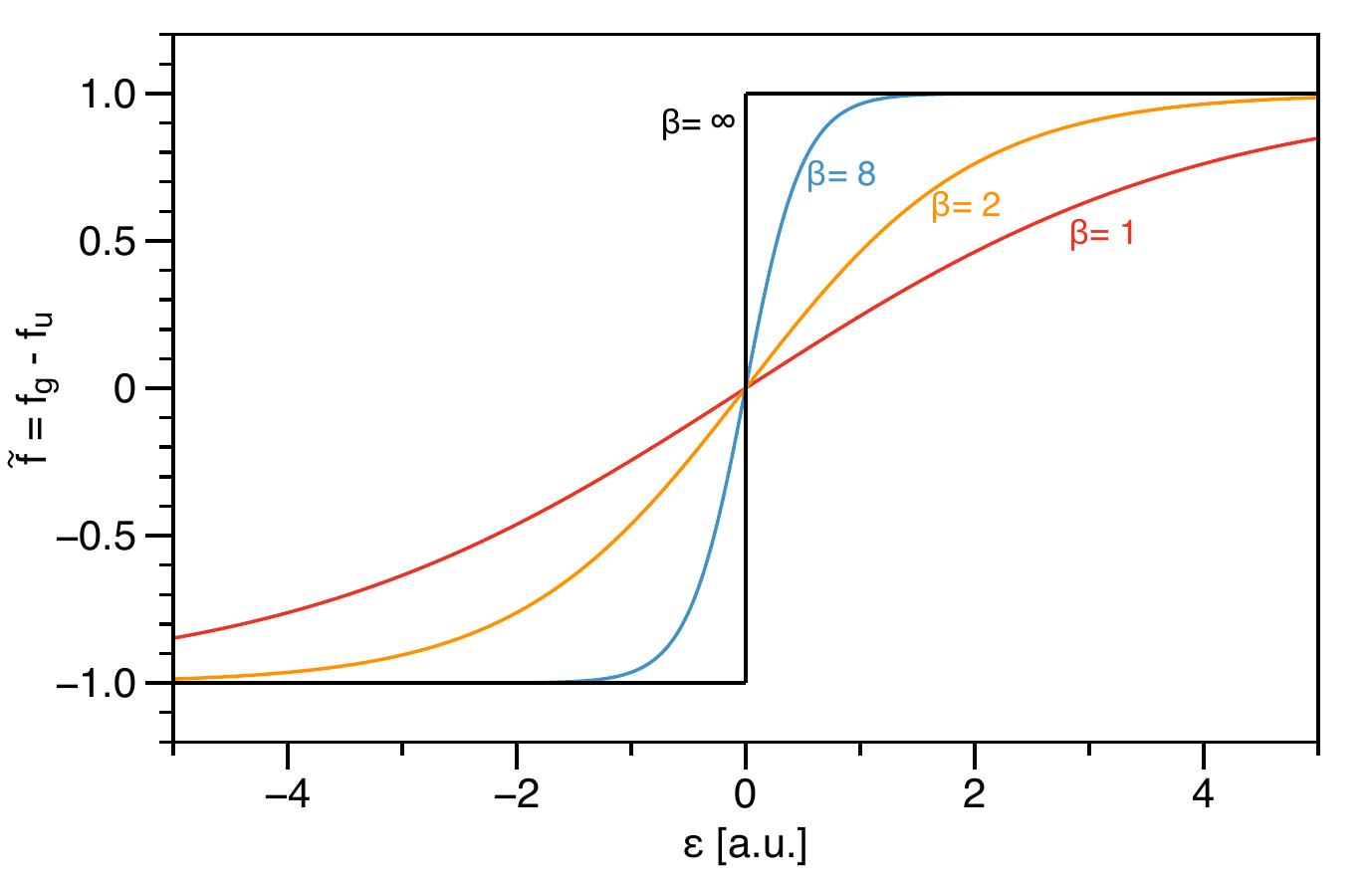}
  \caption{The difference between the (natural) occupation numbers as a function of the gap for several temperatures.}
  \label{fig:gapDifOcc}
\end{figure}

The 1RDM Green's function is defined in a similar manner as the KS Green's function, except that we do not  constrain the effective potential to be local anymore. Allowing for non-local potential, gives more flexibility, but as the potential should retain the symmetry of the system, the non-local 1RDM potential remains diagonal in the symmetry adapted basis of our model system. As the trace of the potential is fixed by the particle number, the non-local potential can only adjust the gap $\epsilon$ in the following manner
\setcounter{equation}{\value{potSave}}
\begin{subequations}
\addtocounter{equation}{2}
\begin{equation}
\vec{v}^{\text{1RDM}}_s(\epsilon)
= \vec{v}^{\text{KS}}_s + \frac{\epsilon + 2t}{2}\begin{pmatrix*}[r] -1 & \;0\; \\ 0 & \;1\; \end{pmatrix*} .
\end{equation}
\end{subequations}%
\setcounter{equation}{\value{eqSave}}%
So in the 1RDM setting, we have one variable to optimise: the gap $\epsilon$. At finite inverse temperature, the relation between the gap and the difference in occupation numbers is one-to-one and explicitly given by
\begin{equation}
\tilde{f} \isDefinedAs f_g - f_u = \tanh(\beta\epsilon/4) .
\end{equation}
In the zero-temperature limit ($\beta \to \infty$), this function collapses into a step function as illustrated in Fig.~\ref{fig:gapDifOcc}. This demonstrates explicitly that the occupation numbers can only be fractional at zero temperature, if the gap closes~\cite{Gilbert1975, Pernal2005, RequistPankratov2008, GiesbertzBaerends2010}. In particular, we have
\begin{align}\label{eq:occDif}
\tilde{f} = \begin{cases}
1		&\text{if $\epsilon > 0$} \\
[-1,1]		&\text{if $\epsilon = 0$} \\
-1		&\text{if $\epsilon < 0$}.
\end{cases}
\end{align}
As we loose the one-to-one relation between the gap and the occupation number difference in the zero-temperature limit, we will need to optimise over the separate pieces of the $\beta = \infty$ curve: $\{\epsilon \leq 0, \tilde{f} = -1\}$, $\{\epsilon = 0, -1 \leq \tilde{f} \leq 1\}$ and $\{\epsilon \geq 0, \tilde{f} = 1\}$.

\subsection{Considerations for the general case}
For simplicity we have limited the discussion on possible input Green's functions to the simple two-orbital model for H$_2$, but all three possibilities can readily be used in more general settings. The simplest case is the use of the HF Green's function, since one only needs to do a (restricted) Hartree--Fock calculation and insert the resulting Green's function into either the Klein~\eqref{eq:KleinFunctional} or Luttinger--Ward functional~\eqref{eq:LWfunctional} with some appropriate approximation for $\Phi[\vec{G}]$. As we will use exclusively the GW approximation for $\Phi$ in this work, we will denote these calculations by K-GW@HF and LW-GW@HF respectively.

The insertion of Kohn--Sham type Green's function results in some variational freedom, as the Kohn--Sham type Green's function encompasses all non-interacting Green's functions which can be generated via a local potential, i.e.\ potentials diagonal in spatial representation or site basis. So in a general setting, we can use the freedom of the local potential optimise the energy expression which results from inserting the KS type Green's function into the Klein or LW functional. Within the GW approximation, we will denote such calculations as K-GW@KS and LW-GW@KS respectively. As discussed before in Sec.~\ref{sec:KSgreen}, the symmetry constraints in the limited two-orbital model (homogeneous density) effectively leaves no variational freedom in the KS potential. This is clearly an artefact of our limited setting.

The space of non-interacting Green's functions to be searched over can be enlarged by allowing the potential to be non-local, i.e.\ we only require the potential to be hermitian ($\vec{v}_s^{\dagger} = \vec{v}_s$). This class of potentials is exactly the type of potentials used in 1RDM functional theory. The insertion of these more general non-interacting Green's functions in the energy functionals within the GW approximation and subsequent optimisation, will therefore be denoted as K-GW@1RDM and LW-GW@1RDM respectively.

\section{Variational calculations}
\label{sec:variational}

In this section we consider the evaluation of the LW and Klein functionals
for various input Green's functions.

To benchmark the results we have also solved
the Dyson equation in the GW approximation self-consistently to obtain the true
stationary Green's function of these functionals. The code has been implemented for spin-independent interactions, so we only have these results for the spin-independent interaction, $\bar{U}=U$. For bond distances larger than $6.4$ Bohr, we had difficulties converging the results. We therefore only include the results up to $6.4$ Bohr, as only those are useful for comparison.

\subsection{Klein functional}
\label{sec:Klein}

\begin{figure}[t]
  \includegraphics[width=\columnwidth]{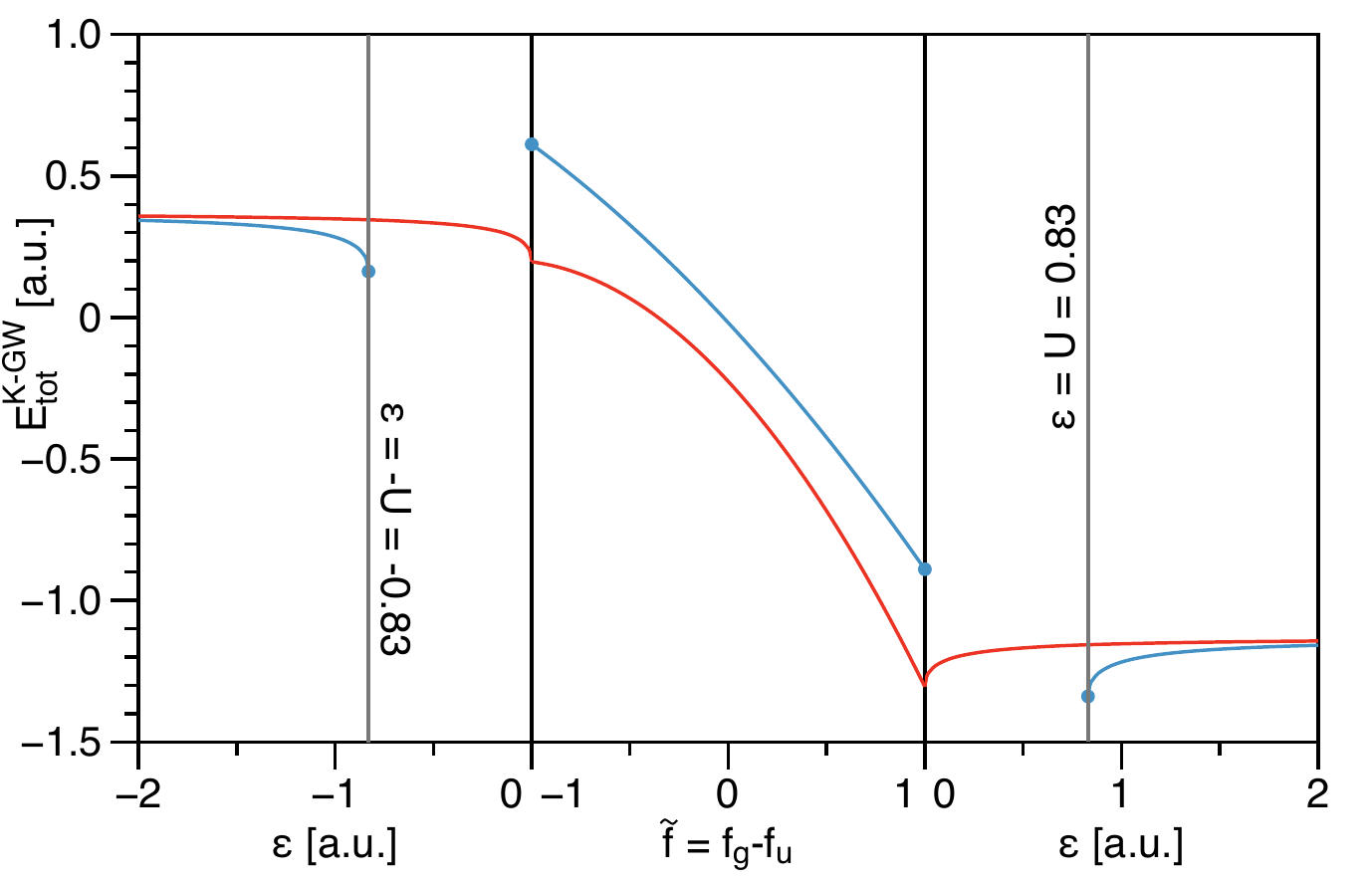}
  \caption{The total GW energy as a function of the gap and occupation number difference at $R_{\text{H--H}} = 1.4$ Bohr. Including the self-interaction term, $\bar{U} = U$, yields a continuous dependence (red), whereas excluding the self-interaction term, $\bar{U} = 0$, results in a discontinuous energy dependence (blue) due to negative numbers in the square root in~\eqref{eq:GWcEnergy}.}
  \label{fig:GWscan}
\end{figure}

The Hartree and exchange part of the $\Phi$ functional were already evaluated in the previous section (Hartree--Fock), so we only need to evaluate the correlation part. The correlation part of the $\Phi$-functional in the GW approximation can be written as
\begin{align}
-\Phi^{\text{GW}}_{\text{c}} &=
\frac{1}{4}
\vcenter{\hbox{%
\begin{tikzpicture}[node distance = 1]
  \node (c2) {};
  \node (d2) [vertex, below = 0.4 of c2] {};
  \node (e2) [vertex, above = 0.4 of c2] {};
  \node (f2) [vertex, right = of d2] {};
  \node (g2) [vertex, right = of e2] {};
  \draw [dressed fermion] (d2) to [bend left=45] (e2);
  \draw [dressed fermion] (e2) to [bend left=45] (d2);
  \draw [bare photon] (f2) -- (d2);
  \draw [bare photon] (e2) -- (g2);
  \draw [dressed fermion] (f2) to [bend left=45] (g2);
  \draw [dressed fermion] (g2) to [bend left=45] (f2);
\end{tikzpicture}%
}}
+ \frac{1}{6}
\vcenter{\hbox{%
\begin{tikzpicture}[node distance = 1]  
  \node (a3) [vertex] {};
  \node (b3) [vertex, above right = 0.75 and 0.5 of a3] {};
  \node (c3) [vertex, right = 0.71 of b3] {};
  \node (d3) [vertex, below right = 0.75 and 0.5 of c3] {};
  \node (e3) [vertex, below left = 0.5 and 0.4 of d3] {};
  \node (f3) [vertex, below right = 0.5 and 0.4 of a3] {};
  \draw [dressed fermion] (a3) to [bend left=45] (b3);
  \draw [dressed fermion] (b3) to [bend left=45] (a3);
  \draw [bare photon] (b3) -- (c3);
  \draw [dressed fermion] (c3) to [bend left=45] (d3);
  \draw [dressed fermion] (d3) to [bend left=45] (c3);
  \draw [bare photon] (d3) -- (e3);
  \draw [dressed fermion] (e3) to [bend left=45] (f3);
  \draw [dressed fermion] (f3) to [bend left=45] (e3);
  \draw [bare photon] (f3) -- (a3);
\end{tikzpicture}%
}} + \dotsb \notag \\
&= \sum_{n=2}^{\infty}\frac{1}{2n}\bigl(\bar{\vec{w}}\vec{P}\bigr)^n \notag \\
&= -\half\Trace\bigl\{\ln\bigl(\vec{1} - \bar{\vec{w}}\vec{P}\bigr) + \bar{\vec{w}}\vec{P}\bigr\} ,
\end{align}
where the trace $\Trace$ is both over matrix entries as well as the Matsubara time\slash{}frequencies.%
\footnote{It is tempting to remove the $\bar{\vec{w}}\vec{P}$ term from the trace, since it seems we would include exchange immediately. However, the time limits in this integral do not correspond to the ones for exchange, so this does not work unfortunately.}
The bar over the interaction indicates that its indices are in chemical ordening, $\bar{w}_{ijkl} \isDefinedAs [il \vert jk] = w_{iklj}$, and the polarization bubble is defined as
\begin{align}
P_{abb'a'}(\omega) &\isDefinedAs \im
\vcenter{\hbox{%
\begin{tikzpicture}[node distance=1]
  \node (a) [vertex] {};
  \node (b) [vertex, right= of a] {};
  \draw [dressed fermion] (b) to [bend left=45] (a);
  \draw [dressed fermion] (a) to [bend left=45] (b);
  \node [anchor=mid, below = 0.15 of a.center] {$b$};
  \node [anchor=mid, below = 0.15 of b.center] {$b'$};
  \node [anchor=mid, above = 0.15 of a.center] {$a$};
  \node [anchor=mid, above = 0.15 of b.center] {$a'$};
  \node at ($ (a)!0.5!(b) $) {$\omega$};
\end{tikzpicture}%
}} \notag \\*
&\hphantom{:}= \frac{1}{\beta}\sum_mG_{a'a}(\omega_m)G_{bb'}(\omega + \omega_m).
\end{align}
in frequency space.
Using that $\trace\bigl\{\ln(\vec{M})\bigr\} = \ln\abs[\big]{\vec{M}}$, i.e.\ the logarithm of the determinant, the expression for the correlation part of the $\Phi$-functional in the GW approximation can be further simplified to
\begin{multline}
\Phi^{\text{GW}}_{\text{c}}
= \half\sum_p\e^{\eta\omega_p}\bigl(\ln\abs[\big]{\vec{1} - \bar{\vec{w}}\vec{P}(\omega_p)} \\*
{} + \trace\{\bar{\vec{w}}\vec{P}(\omega_p)\}\bigr),
\end{multline}
where the trace, $\trace$, is now only over matrix entries.
For our simple non-interacting Green's function, the determinant can be worked out as (see supplement)
\begin{multline}
\begin{vmatrix} \vec{1} - \bar{\vec{w}}\vec{P}_s(\omega) \end{vmatrix}
= \left(1 - \frac{2\epsilon(u + v)\tilde{f}}{\omega^2 - \epsilon^2}\right) \\
{} \times \left(1 - \frac{2\epsilon(u - v)\tilde{f}}{\omega^2 - \epsilon^2}\right),
\end{multline}
where $\epsilon \isDefinedAs \epsilon_u - \epsilon_g$ denotes the gap and we introduced the short-hand notations $u \isDefinedAs \half\bigl(\bar{U} - w\bigr)$ and $v \isDefinedAs \half\bigl(U - w\bigr)$. Evaluating the remaining sum over the bosonic frequencies (see supplement) yields
\begin{align}\label{eq:PhiGWc}
\Phi^{\text{GW}}_{\text{c},s}
={}& \ln\biggl(\sinh\Bigl(\frac{\beta\zeta_+}{2}\Bigr)\biggr) +
\ln\biggl(\sinh\Bigl(\frac{\beta\zeta_-}{2}\Bigr)\biggr) \\*
&{} - 2\ln\biggl(\sinh\Bigl(\frac{\beta\abs{\epsilon}}{2}\Bigr)\biggr) -
\beta\frac{u}{2}\bigl(\tilde{f}^2 + \tilde{n}(2 - \tilde{n})\bigr) , \notag
\end{align}
where
\begin{equation}
\zeta_{\pm} \isDefinedAs \sqrt{\epsilon^2 + 2\epsilon(u \pm v)\tilde{f}} .
\end{equation}
Taking the $T \to 0$ limit, we find that the Klein functional gives the following approximation for the correlation energy for $\tilde{n} = 1$
\begin{equation}\label{eq:GWcEnergy}
E^{\text{K-GW}}_{\text{c}}\! = \lim_{\mathclap{\beta \to \infty}}\,\frac{1}{\beta}\Phi^{\text{GW}}_{\text{c}}
= \frac{\zeta_+}{2} + \frac{\zeta_-}{2} - \abs{\epsilon} - \frac{u(1 + \tilde{f}^2)}{2} .
\end{equation}
Since the K-GW correlation energy~\eqref{eq:GWcEnergy} depends explicitly on the self-interaction term $\bar{U}$, we should investigate different values. As mentioned in the introduction, we will investigate
a spin-independent interaction ($\bar{U} = U$) and the explicitly self-interaction free model ($\bar{U}=0$).
The total K-GW energy is plotted as a function of the gap\slash{}occupation numbers for these two cases in Fig.~\ref{fig:GWscan} for $R_{\text{H--H}} = 1.4$ Bohr 
along the 3 segments of the $\beta=\infty$-path as depicted in Fig.~\ref{fig:gapDifOcc}.
In the case of a spin-independent interaction, the K-GW energy can be calculated for all values of the gap as is shown by the continuous line. If the self-interaction term is set to zero, $\bar{U} = 0$, small values of the gap yield an imaginary part to the energy, since we have $u - v = -U/2 <0$ in the second square root $\zeta_-$ of the K-GW correlation energy~\eqref{eq:GWcEnergy}. Only when $\abs{\epsilon} \geq U$ or $\epsilon = 0$, we obtain a real value for the K-GW energy as is apparent from the discontinuous curve in Fig.~\ref{fig:GWscan}.

The relative positioning of the minima of the total GW energy as depicted in Fig.~\ref{fig:GWscan} turns out to be the same at all bond distances. The minimum is always located at a fully occupied gerade orbital $\tilde{f} = 1$ for both choices of $\bar{U}$. In the case of a spin-independent interaction, $\bar{U} = U$, the optimal gap is always located at $\epsilon = 0$. Excluding self-interaction term, $\bar{U} = 0$, always yields the global minimum at the smallest strictly positive gap, $\epsilon = U$.

\begin{figure}[t]
  \includegraphics[width=\columnwidth]{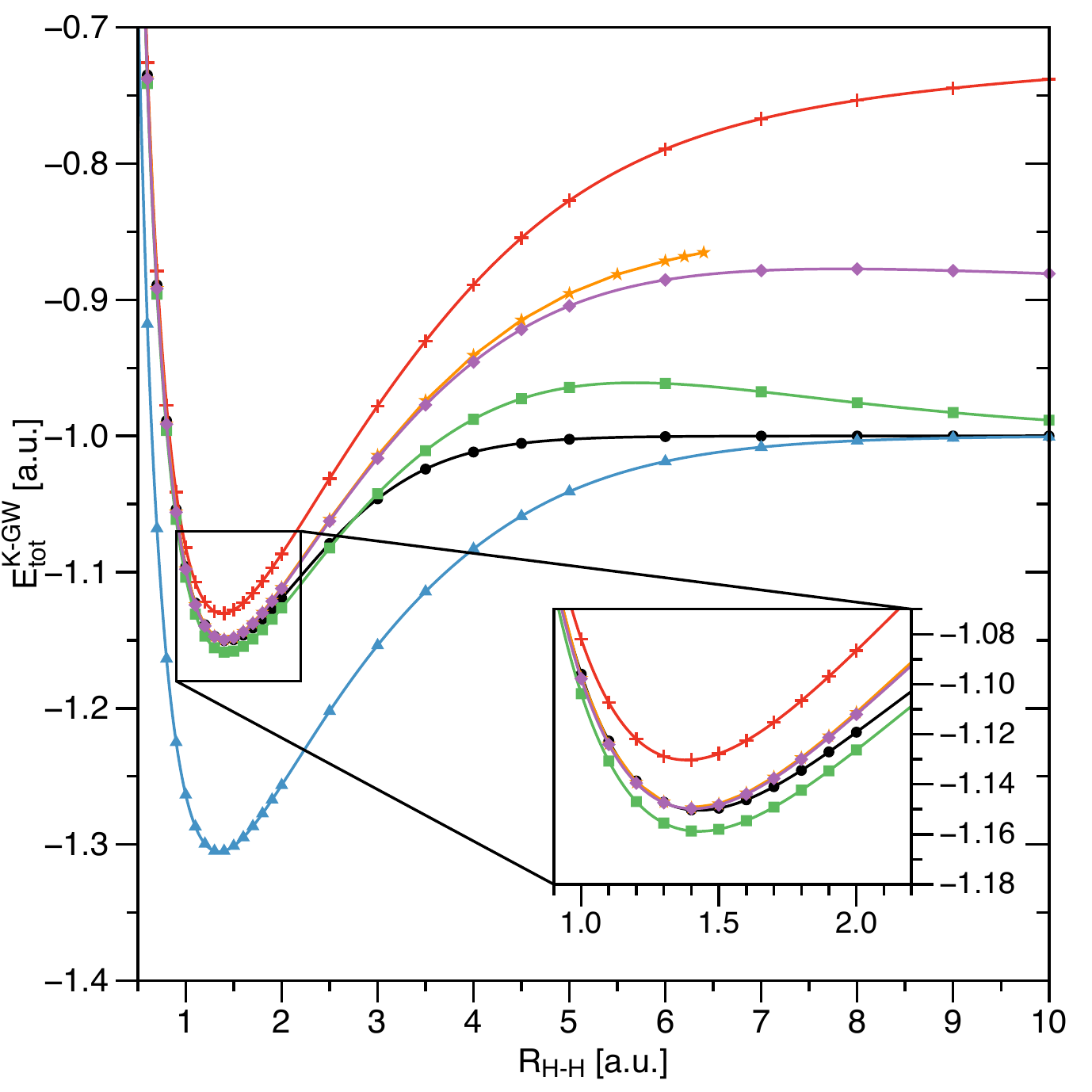}
  \caption{The ground state energy calculated with the Klein-GW functional compared to the exact Hubbard results (black circles), HF (red crosses) and the self-consistent GW energy (orange stars). The self-interaction term is included: $\bar{U} = U$. The following three different versions are shown: the 1RDM version (blue triangles), the DFT version (green squares) and the HF version (purple diamonds).}
  \label{fig:GW-UbU}
\end{figure}

Different methods to use the K-GW energy expression~\eqref{eq:GWcEnergy} in the spin-independent interaction case are compared to the exact ground state in Fig.~\ref{fig:GW-UbU}. The 1RDM version of the K-GW functional (K-GW@1RDM) is consistently too low, but becomes exact in the dissociation limit. Restricting the potential to be local (K-GW@KS) raises the total energy and it becomes quite accurate around the equilibrium distance, though still slightly too low. In the dissociation regime, the upshift is too high and creates the infamous RPA bump. Since the KS gap also vanishes in the dissociation limit, the correct dissociation limit is retained~\cite{FuchsNiquetGonze2005,DahlenLeeuwenBarth2006}. Alternatively, one could use the HF gap as input (K-GW@HF). The HF gap is larger than the KS gap due to the additional non-local exchange in the potential~\eqref{eq:HFgap}. This leads to a higher total energy, since the K-GW correlation energy increases monotonically in $\epsilon$ for positive gaps as can be seen from~\eqref{eq:GWcEnergy}. The K-GW@HF energy gives an improvement over the KS gap result around the equilibrium distance, in the sense that the total energy agrees even better with the exact result (see Fig.~\ref{fig:GW-UbU}). Upon dissociation however, the artificial bump is raised to even higher values, leading to a worse performance than the KS gap. Though not so clear from Fig.~\ref{fig:GW-UbU}, the dissociation limit remains correct, since also the HF gap goes to zero when $R_{\text{H--H}} \to \infty$. Albeit, the closure rate is much slower. Due to the additional exchange contribution to the gap, the HF gap only closes as $1/R_{\text{H--H}}$, whereas the KS gap closes exponentially.

\begin{figure}[t]
  \includegraphics[width=\columnwidth]{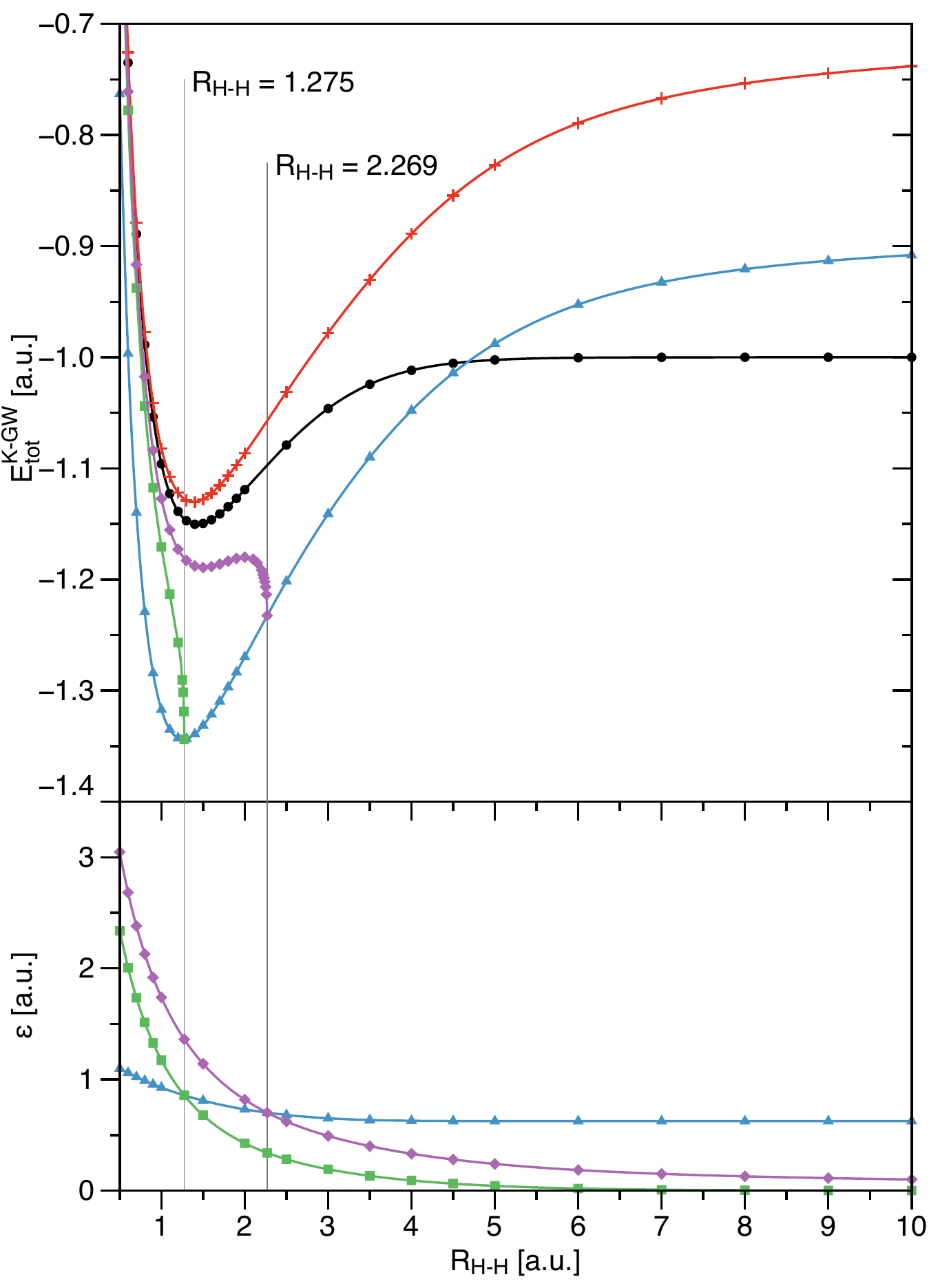}
  \caption{The upper panel is the same as in Fig.~\ref{fig:GW-UbU} but now without the self-interaction term: $\bar{U} = 0$. In the lower panel the value of the gap is shown which went into the GW correlation energy expression~\eqref{eq:GWcEnergy}.}
  \label{fig:GW-Ub0}
\end{figure}

Since the GW approximation is not self-interaction free (depends on $\bar{U}$), one would expect that the results would improve if the self-interaction term would be explicitly excluded from the Hamiltonian. The results for the total energy for $\bar{U} = 0$ are shown in the upper panel of Fig.~\ref{fig:GW-Ub0}. In fact, elimination of the self-interaction does actually not lead to any improvement at all. Using the K-GW functional as a 1RDM functional leads to even lower energies around the equilibrium distance and upon dissociation the energy becomes higher than the exact results. Even the dissociation limit is not correct anymore, since the gap needs to remain finite, $\epsilon = U$, to prevent the energy from having a complex part, as mentioned before in connection with Fig.~\ref{fig:GWscan}. The results are even more disastrous when the KS or HF gaps are used. Since both gaps decrease when stretching the bond, there is always a point from where they become smaller than $U$ and the K-GW correlation energy becomes complex. This is illustrated in the lower panel in Fig.~\ref{fig:GW-Ub0}, where the different gaps of each calculation are plotted. Since the 1RDM gap is always the minimum gap, the failure of the K-GW functional with the KS\slash{}HF gap exactly occurs when it crosses the 1RDM gap $\epsilon = U$. At these points, the K-GW energy from the KS\slash{}HF gap becomes exactly equal to the K-GW@1RDM functional.

\subsection{Luttinger--Ward functional}
\label{sec:LW}
In an attempt to improve the results, one can use the Luttinger--Ward (LW) functional instead of the Klein functional, as the LW functional has superior variational properties~\cite{DahlenBarth2004,DahlenLeeuwenBarth2006}. The superior variational properties come at the cost of a more complicated functional which requires the evaluation of the following additional terms
\begin{align}\label{eq:LWcor}
\Omega_{\text{LW}} - \Omega_{\text{K}}
= -\frac{1}{\beta}\Trace\bigl\{\vec{\tilde{\Sigma}}\vec{G} + \ln\bigl(\vec{1} - \vec{G}_s\vec{\tilde{\Sigma}}\bigr)\bigr\} ,
\end{align}
where the modified self-energy $\vec{\tilde{\Sigma}}$ was defined in~\eqref{eq:modSigmaDef}.

Since the HF part has already been evaluated before, cf.~\eqref{eq:SigmaHF} and~\eqref{eq:HFpotential}, we only need the correlation part of the self-energy in the GW approximation evaluated at $\vec{G}_s$.
The evaluation is somewhat simplified by the fact that the self-energy is also diagonal in the symmetry-adapted basis. Its diagonal elements can be evaluated to be
\begin{subequations}\label{eq:SigmaGW}
\begin{align}
\Sigma^{\text{GW}}_{\text{c},s;gg}[\vec{G}_s](\omega)
&= R_+(\omega - \epsilon^{\text{M}}_u) , \\
\Sigma^{\text{GW}}_{\text{c},s;uu}[\vec{G}_s](\omega)
&= R_-(\omega - \epsilon^{\text{M}}_g) ,
\end{align}
\end{subequations}
where we introduced
\begin{align}
R_{\pm}(\omega) \isDefinedAs \epsilon\tilde{f}\biggl[&
\frac{(u+v)^2}{2\zeta_+}\frac{\omega}{\omega^2 - \zeta_+^2}\coth\Bigl(\frac{\beta\zeta_+}{2}\Bigr)
\notag \\*
{} + {}& \frac{(u-v)^2}{2\zeta_-}\frac{\omega}{\omega^2 - \zeta_-^2}\coth\Bigl(\frac{\beta\zeta_-}{2}\Bigr) \\*
{} + {}&\frac{(u^2+v^2)(\omega^2 - \epsilon^2) -2\epsilon\tilde{f}(u^2 - v^2)}
{(\omega^2 - \zeta_+^2)(\omega^2 - \zeta_-^2)} \notag \\*
&\qquad\qquad\qquad\qquad\qquad {} \times \bigl(1 - \tilde{n} \pm \tilde{f}\bigr)\biggr] . \notag
\end{align}
For $\tilde{n} = f_g + f_u = 1$, $R_{\pm}$ satisfies $R_+(-\omega) = -R_-(\omega)$. As in that case also $\epsilon_g^M = -\epsilon_u^M$, the components of the self-energy are simply related as
\begin{align}\label{eq:SigmaGWsym}
\Sigma^{\text{GW}}_{\text{c},s;gg}[\vec{G}_s](-\omega)
&= -\Sigma^{\text{GW}}_{\text{c},s;uu}[\vec{G}_s](\omega) &
&\text{if $\tilde{n} = 1$.}
\end{align}
This relation is convenient, as it implies that the number of particles is not affected by the correlation part for $\tilde{n} = 1$ (see Appendix~\ref{ap:GWchemPot}), proving the validity of the constant in (trace of) the potentials used to generate the input Green's functions~\eqref{eq:pots}.

\begin{figure}[t]
  \includegraphics[width=\columnwidth]{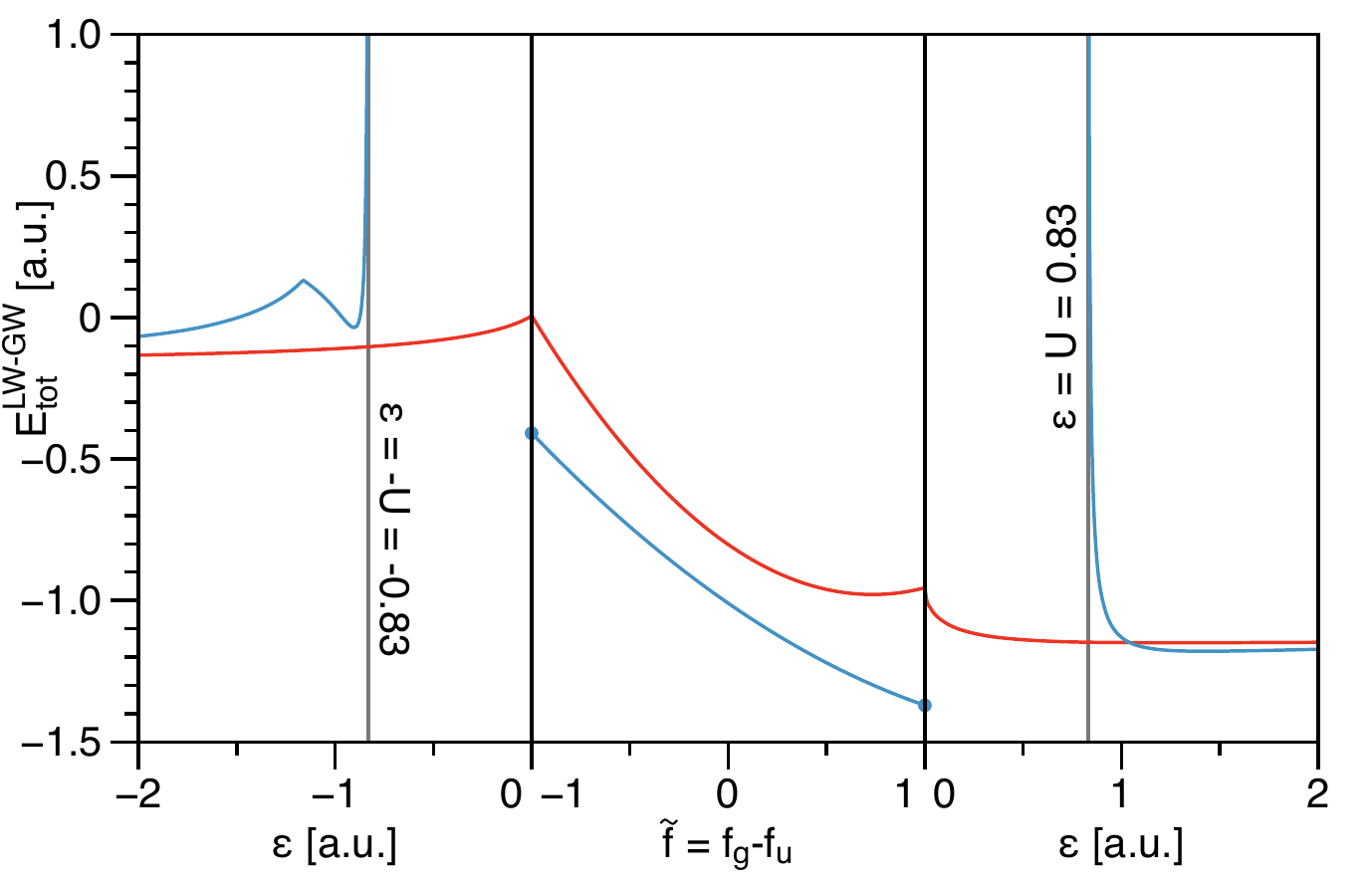}
  \caption{The total LW-GW energy as a function of the gap and occupation number difference at $R_{\text{H--H}} = 1.4$ Bohr. Including the self-interaction term, $\bar{U} = U$, yields a continuous dependence (red), whereas excluding the self-interaction term, $\bar{U} = 0$, results in a discontinuous energy dependence (blue) due to negative numbers in the square root in~\eqref{eq:GWcEnergy}.}
  \label{fig:LWGWscanEq}
  \includegraphics[width=\columnwidth]{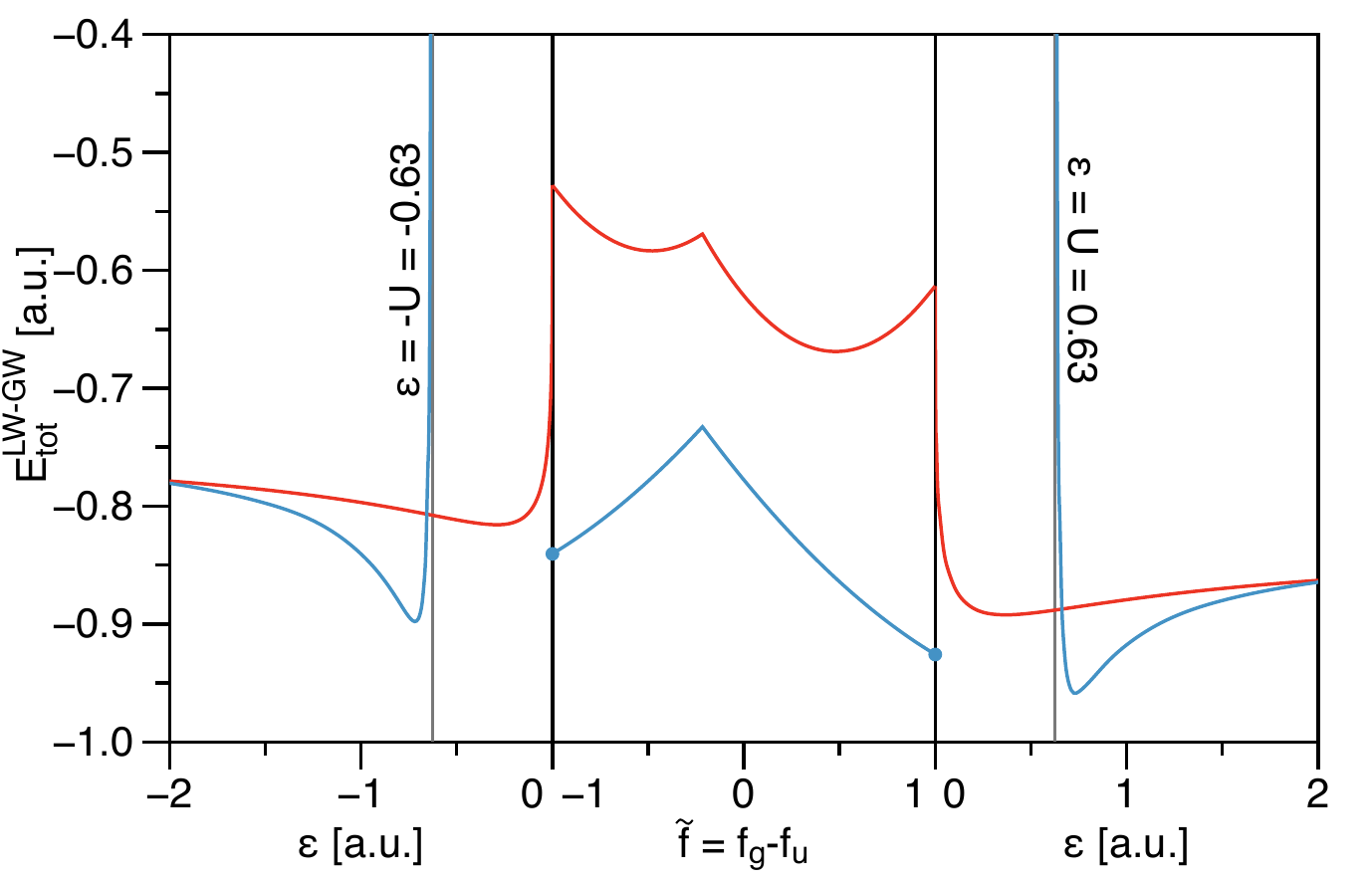}
  \caption{The total LW-GW energy as a function of the gap and occupation number difference at $R_{\text{H--H}} = 5$ Bohr. Again, the continuous (red) line corresponds to the spin-dependent interaction, $\bar{U} = U$, and the the discontinuous one (blue) the self-interaction free interaction $\bar{U} = 0$.}
  \label{fig:LWGWscan5}
\end{figure}

The linear term of the LW correction in~\eqref{eq:LWcor} can now be evaluated to be
\begin{multline}\label{eq:LWlinGW}
\Trace\bigl\{\vec{\Sigma}^{\text{GW}}_{\text{c},s}\vec{G}_s\bigr\}
= \frac{\beta\epsilon\tilde{f}(u+v)}{\zeta_+}\coth\Bigl(\frac{\beta\zeta_+}{2}\Bigr) \\*
{} + \frac{\beta\epsilon\tilde{f}(u - v)}{\zeta_-}\coth\Bigl(\frac{\beta\zeta_-}{2}\Bigr) -
\beta\bigl(\tilde{n}(2 - \tilde{n}) + \tilde{f}^2\bigr)u .
\end{multline}
Unfortunately, the logarithmic term in~\eqref{eq:LWcor} cannot be evaluated analytically. The integrand decays as $1/\omega^2$, so numerical evaluation is not a problem. In the gapless limit, $\epsilon \to 0$, however, an analytical solution is in reach. In the limit of vanishing gap, the self-energy simplifies considerably to
\begin{equation}
\lim_{\epsilon \to 0}\Sigma^{\text{GW}}_{\text{c},s;gg/uu}(\omega + \epsilon^{\text{M}}_{u/g})
= \frac{u}{\beta\omega} .
\end{equation}
In the $\beta \to \infty$ limit its contribution will therefore vanish, so only the contribution of the correlation potential in the logarithmic term remains. Evaluation of the logarithmic term with only the correlation potential is rather straightforward and gives
\begin{equation}
\lim_{\mathclap{\beta\to\infty}}\,\frac{1}{\beta}
\Trace\bigl\{\ln\bigl(\vec{1} +\vec{v}^{\text{c}}\vec{G}_s\bigr)\bigr\}
= 2\abs{v^{\text{c}}_{gg}} .
\end{equation}
where
\begin{equation}
\vec{v}^c \isDefinedAs \vec{v}_s - \vec{v}_{\text{ext}} - \vec{v}^{\text{HF}} .
\end{equation}
The values for the total energies evaluated with the LW functional in the GW approximation are plotted as a function of the gap and the occupation number difference in Fig.~\ref{fig:LWGWscanEq} for a bond distance of $R_{\text{H--H}} = 1.4$ Bohr. In the case of a spin-independent interaction, $\bar{U} = U$, the total energy can be calculated for any gap and the minimum is now located at a positive gap. In the case of the self-interaction free interaction, $\bar{U} = 0$, the energy is only real when the gap is either sufficiently large or zero. This is exactly the same situation as described before for the Klein functional. For short bond distances the minimum is located at $\tilde{f} = 1$ and $\epsilon = 0$, but at larger bond distances the minimum shifts to a positive gap. This relocation of the minimum is demonstrated in Fig.~\ref{fig:LWGWscan5} for $R_{\text{H--H}} = 5.0$ Bohr, and occurs around $R_{\text{H--H}} = 4.25$ Bohr. The relocation of the minimum only occurs for self-interaction free case and not for the spin-independent interaction case.

\begin{figure}[t]
  \includegraphics[width=\columnwidth]{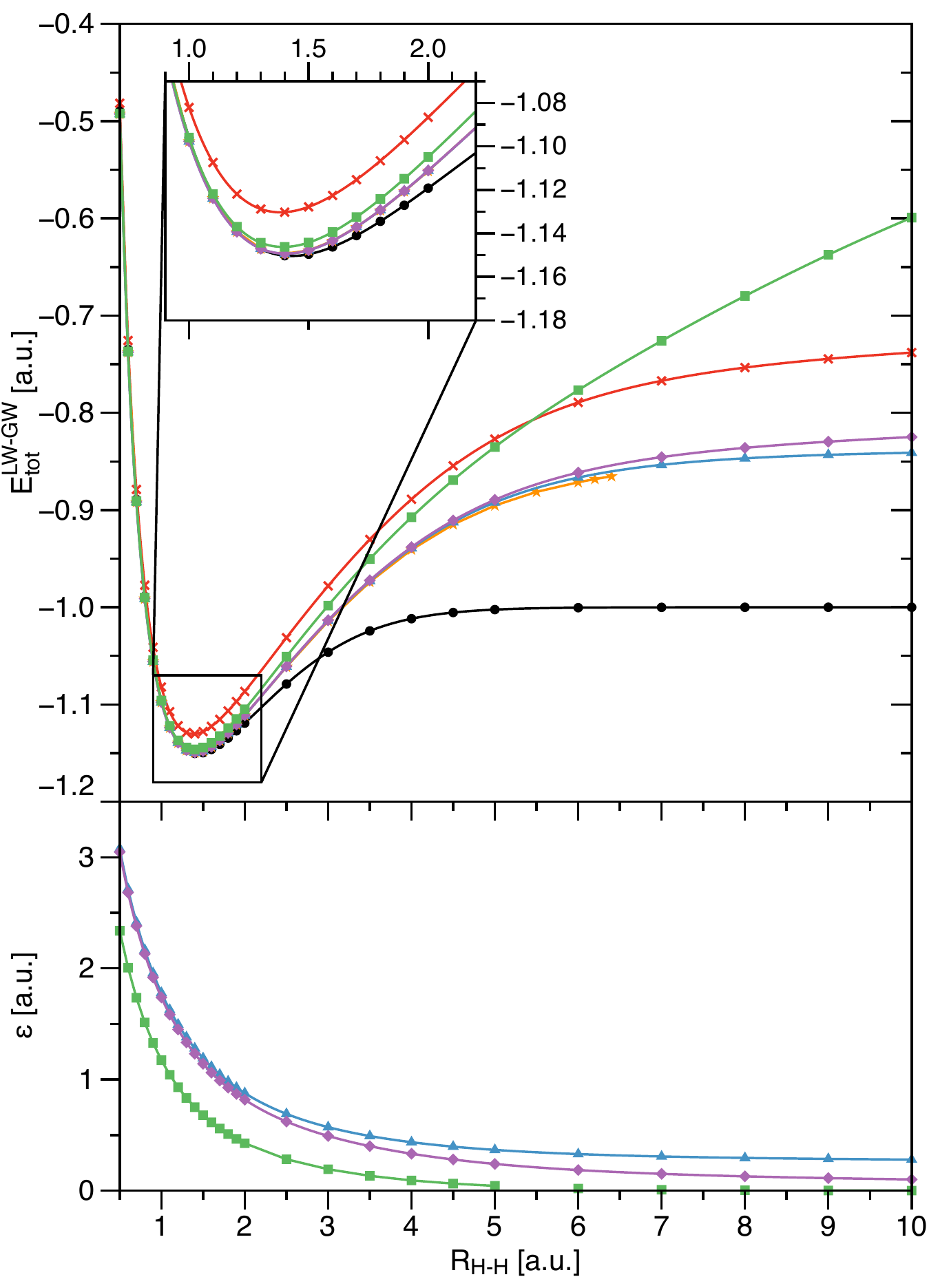}
  \caption{Upper panel: the ground state energy calculated with the Luttinger--Ward-GW functional compared to the exact Hubbard results (black circles), HF (red crosses) and the
  self-selfconsistent GW energy (orange stars). The self-interaction term is included: $\bar{U} = U$. The following three different versions are shown: the 1RDM version (blue triangles), the DFT version (green squares) and the HF version (purple diamonds). In the lower panel the value of the gap is shown which went into the LW-GW correlation energy expression.}
  \label{fig:LWGW-UbU}
\end{figure}

In the top panel of Fig.~\ref{fig:LWGW-UbU} we show the results for the total energies from the Luttinger--Ward functional for the spin-free Hamiltonian, $\bar{U} = U$. Compared to the Klein functional we see a huge improvement of the total energy when the LW functional is used as a 1RDM functional, but unfortunately, the dissociation limit is not correct anymore. As a matter of fact, the optimised 1RDM results are almost identical to the results when the HF Green's function is used as input. Only in the dissociation limit the optimisation of the non-local potential yields a somewhat lower result. This trend is corroborated by comparing the HF gap and the 1RDM gap visualised in the lower panel of Fig.~\ref{fig:LWGW-UbU}. At equilibrium distance the gaps are nearly identical and they start to deviate somewhat when the interatomic distance is increased. Somewhat surprisingly, increasing the gap slightly from its HF value yields a lower energy.

We have also included the results when the KS Green's function is used as input for the LW functional. The KS Green's function now gives the worst result of all Greens functions, which is somewhat surprising, as it performed so well in the Klein functional. The results for the Klein functional are mainly due to error cancelations between the inflexibility of the KS Green's function and the bad variational properties of the Klein functional. A similar inferior behaviour of the KS Green's function as input for the LW functional has been obtained for the H$_2$ molecule in a large Slater basis~\cite{DahlenLeeuwenBarth2006}.

\begin{figure}[t]
  \includegraphics[width=\columnwidth]{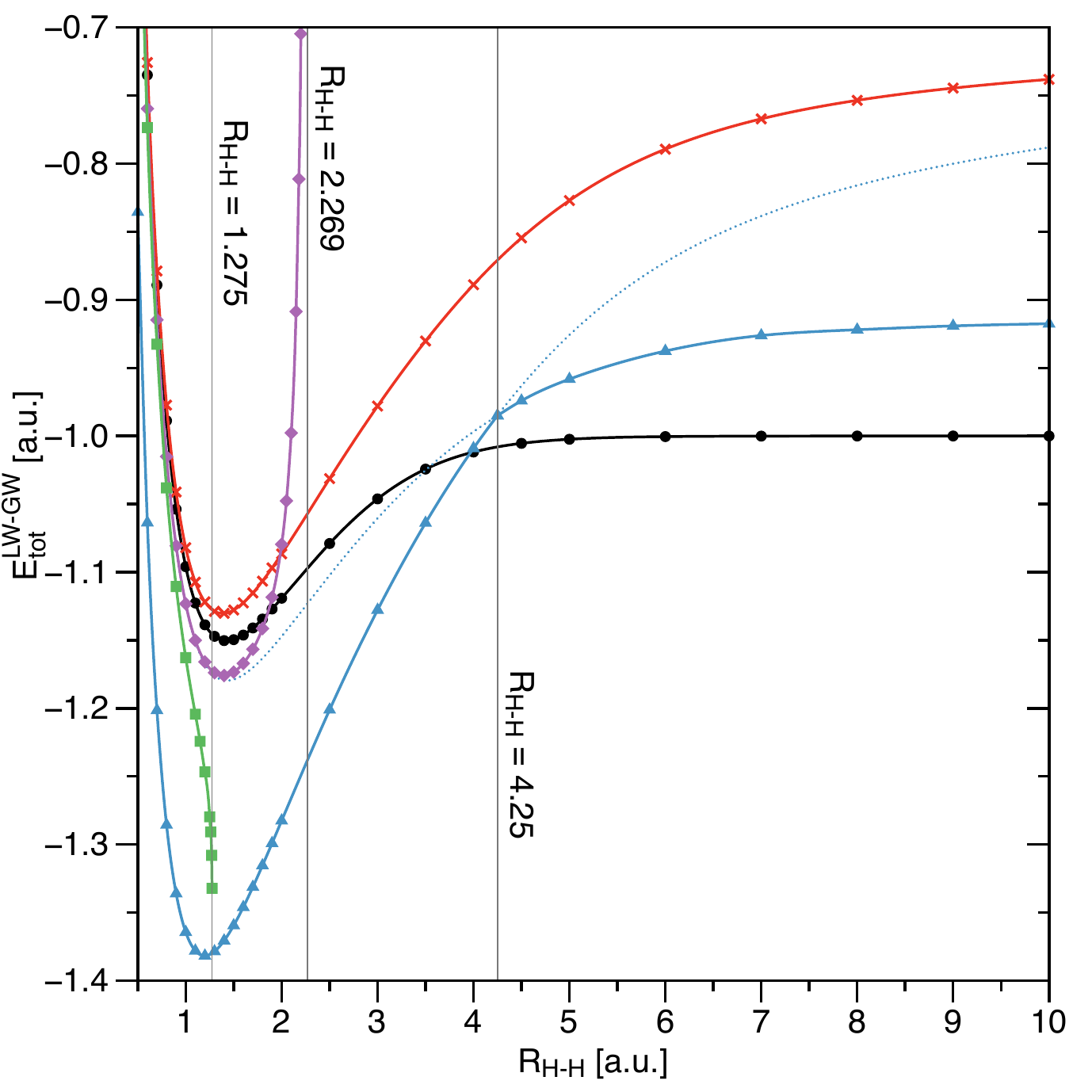}
  \caption{The same as the upper panel in Fig.~\ref{fig:LWGW-UbU}, though now with the self-interaction free interaction, $\bar{U} = 0$. The next higher energy solutions when using the LW as a 1RDM functional are indicated by the (blue) dotted lines.}
  \label{fig:LWGW-Ub0}
\end{figure}

Now let us consider the results in the self-interaction free case, $\bar{U} = 0$. The results for the total energies are depicted in Fig.~\ref{fig:LWGW-Ub0}. Most notable are similar catastrophes for the KS and HF input Green's functions as for the Klein functional, cf.\ Fig.~\ref{fig:GW-Ub0}. When the bond is sufficiently stretched, both the KS and HF Green's function yield from some distance onwards a complex energy, due to the square root in $\zeta_-$. These catastrophes therefore occur at exactly the same bond distances as for the Klein functional. The only difference is that for the HF Green's function the catastrophe for the LW functional occurs in a different manner than for the Klein functional. In the case of the KS functional the behaviour of the two functionals is nearly identical.

When using the full flexibility of a non-local potential to generate a trial Green's function, the LW functional does not perform better than the Klein functional in the $\bar{U} = 0$ case. Around the equilibrium distance the total energy is somewhat lower compared to the Klein functional, so the LW functional yields worse results. Approaching the dissociation limit, $R_{\text{H--H}} = 10.0$ Bohr, the results are somewhat improved. Arguably, the most interesting feature when using the LW as a 1RDM functional is the kink at $R_{\text{H--H}} = 4.25$ Bohr. This is the location where the global minimum jumps from the $\epsilon = 0$ solution to $\epsilon > U$ solution (see Figs.~\ref{fig:LWGWscanEq} and~\ref{fig:LWGWscan5}). When one of these solutions becomes higher in energy, we have continued plotting the solution with a dotted line in Fig.~\ref{fig:LWGW-Ub0}. If now the $\epsilon = 0$ solution would be disregarded altogether, we see that the finite gap solution actually yields a quite reasonable result. We could not think of a proper physical argument to discard the gapless solution, so we do not promote this procedure. All these issues remain a topic for future research.

\section{Conclusion}
\label{sec:conclusion}

In this work we considered two different energy functionals of the 1RDM. These were obtained by restricting the domain of the 
LW and Klein functionals to non-interacting Green's function for a Hamiltonian with a spatially non-local potential.
These functionals were evaluated within the GW approximation for two different cases of self-interaction for an extended Hubbard Hamiltonian
representing a model system for the hydrogen molecule. We compared the binding curves of this system with exact results as well
as with energy functionals with HF and KS input Green's functions. We found that the LW functional outperforms the Klein functional
and gives results very close to the optimal self-consistent GW Green's function which shows that there is room for future improvement of
the method. Especially the GW approximation does not correctly treat the self-interaction correctly and a probable improvement 
can be attained by including T-matrix and ladder diagram expansions to introduce the necessary exchange-type diagrams.
The system that we considered is very rigid due to its low dimensionality and systems in larger basis sets are likely to increase the variational freedom
to attain more accurate results. We further studied the system in the zero-temperature limit which led to discontinuities in the energy landscape.
It is therefore worthwhile to further explore systems at finite temperature to smoothen these discontinuities. 

\begin{acknowledgement}
KJHG thanks B.C.E. Giesbertz for useful suggestions and gratefully acknowledges a VENI grant by the Netherlands Foundation for Research NWO (722.012.013).
\end{acknowledgement}

\appendix
\numberwithin{equation}{section}

\section{Derivation of matrix elements of the two-orbital model for H$_2$}
\label{ap:matElems}

\begin{figure}[t]
  \includegraphics[width=\columnwidth]{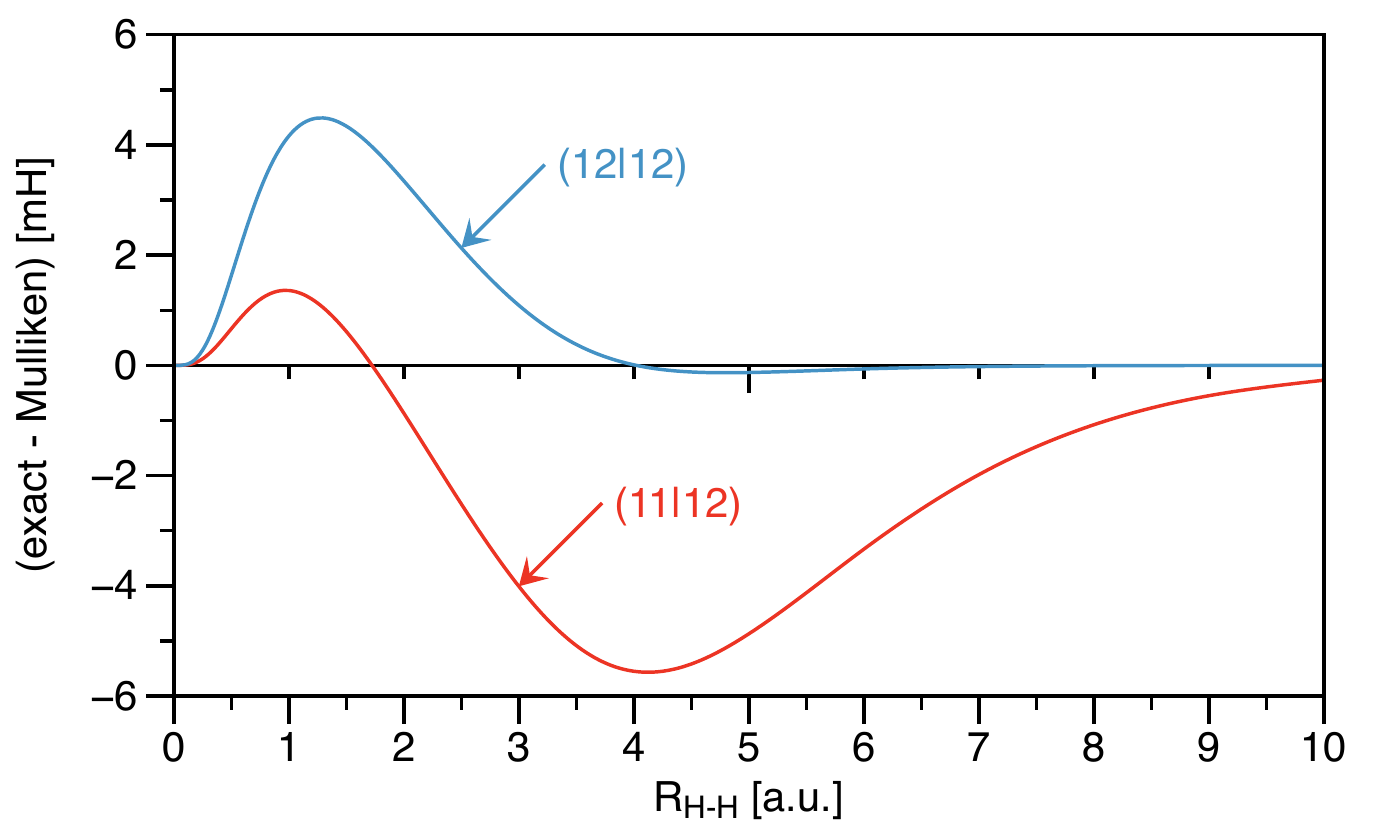}
  \caption{Difference between the exact integrals and the integrals in the Mulliken approximation. Note that the error is in milliHartree.}
  \label{fig:integralsCheck}
\end{figure}

Defining $\rho \isDefinedAs \zeta R$, the one-electron integrals in the non-orthogonal 1s-basis become
\begin{align}
s \isDefinedAs \braket{\chi_1}{\chi_2} &= \left(1 + \rho + \frac{\rho^2}{3}\right)\e^{-\rho} , \notag \\
\brakket{\chi_1}{\hat{h}}{\chi_1} &= \frac{\zeta^2}{2} - \zeta + \frac{\zeta}{\rho}\left(\e^{-2\rho}(1 + \rho) - 1\right) , \\
\brakket{\chi_1}{\hat{h}}{\chi_2} &= \e^{-\rho}\left[\frac{\zeta^2}{2}\left(1 + \rho - \frac{\rho^2}{3}\right) - 2\zeta(1 + \rho)\right] . \notag
\end{align}
The unique two-electron integrals in the non-orthogonal 1s-basis are
\begin{subequations}
\begin{align}
(11 \vert 11) &= \frac{5\zeta}{8} , \\
(11 \vert 22) &= \zeta\left[\frac{1}{\rho} - \e^{-2\rho}\left(\frac{1}{\rho} + \frac{11}{8} + \frac{3\rho}{4} + \frac{\rho^2}{6}\right)\right] , \\
(11 \vert 12) &= \frac{\zeta}{16\rho}\bigl[\e^{-\rho}(5 + 2\rho + 16\rho^2) \notag \\*
&\eqspace\quad
{} - \e^{-3\rho}(5 + 2\rho)\bigr] , \\
\label{eq:1212}
(12 \vert 12) &= \frac{\zeta\e^{-2\rho}}{120\rho}
\Bigl[75\rho - 138\rho^2 - 72\rho^3 - 8\rho^4 \notag \\*
&\eqspace
{} + 16\bigl(\gamma+\ln(\rho)\bigr)\bigl(9 + 18\rho + 15\rho^2 + 6\rho^3 + \rho^4\bigr) \notag \\*
&\eqspace
{} - 32\bigl(9 - 3\rho^2 + \rho^4\bigr)\e^{2\rho}\Ei(-2\rho) \notag \\*
&\eqspace
{} + 16(3 - 3\rho + \rho^2)^2\e^{4\rho}\Ei(-4\rho)\Bigr],
\end{align}
\end{subequations}
where $\gamma \approx 0.5772$ is the Euler--Mascheroni constant and we used the exponential integral
\begin{equation}
\Ei(x) \isDefinedAs -\binteg{y}{-x}{\infty}\frac{\e^{-y}}{y} .
\end{equation}
The formula~\eqref{eq:1212} has been derived using the procedure presented in~\cite{Rosen1931}
and agrees to all digits with the numerical results presented in~\cite{HirschfelderLinnett1950}.
In the calculations we use the Mulliken approximation to the two-electron integrals~\cite{Mulliken1949}
\begin{equation}
(il \vert jk) \approx \frac{S_{il}S_{jk}}{4}\bigl[(ii \vert jj) + (ll \vert jj) + (ii \vert kk) + (ll \vert kk)\bigr] ,
\end{equation}
where $S_{ij} \isDefinedAs \braket{\chi_i}{\chi_j}$ is the overlap.
That means that the following integrals are approximated as
\begin{subequations}
\begin{align}
(11 \vert 12) &\approx \frac{s}{2}\bigl[(11 \vert 11) + (11 \vert 22)\bigr] , \\
(12 \vert 12) &\approx \frac{s^2}{2}\bigl[(11 \vert 11) + (11 \vert 22)\bigr] .
\end{align}
\end{subequations}
The Mulliken approximation has been verified numerically as a function of the bond length combined with the optimized exponent. The error of the Mulliken approximation is shown in Fig.~\ref{fig:integralsCheck} and is of the order of mH in around the equilibrium distance and decreases quickly for the $(12 \vert 12)$ integral upon dissociation.

When transforming to the Löwdin orthogonalised basis we obtain exactly the one-electron elements used in the extended Hubbard model~\eqref{eq:LowdinOneElectron}. The transformed two-electron integrals can be worked out as
\begin{subequations}
\begin{align}
[11 \vert 11] &= \frac{1}{2(1-s^2)^2}\bigl[(2 - s^2)(11\vert11) + s^2(11\vert22) \notag \\*
&\eqspace\qquad\qquad
{} - 4s(11\vert12) + 2s^2(12\vert12)\bigr], \\
[11 \vert 22] &= \frac{1}{2(1-s^2)^2}\bigl[(2 - s^2)(11\vert22) + s^2(11\vert11) \notag \\*
&\eqspace\qquad\qquad
{} - 4s(11\vert12) + 2s^2(12\vert12)\bigr], \\
[11 \vert 12] &= \frac{1}{2(1-s^2)^2}\bigl[2(1 + s^2)(11\vert12) - 2s(12\vert12) \notag \\*
&\eqspace\qquad\qquad
{} - s\bigl((11\vert11) + (11\vert22)\bigr)\bigr], \\
[12 \vert 12] &= \frac{1}{2(1-s^2)^2}\bigl[2(12\vert12) - 4s(11\vert12) \notag \\*
&\eqspace\qquad\qquad
{} + s^2\bigl((11\vert11) + (11\vert22)\bigr)\bigr].
\end{align}
\end{subequations}
Now using the Mulliken approximation, the expressions for the two-electron integrals in the Löwdin orthogonalised basis simply to
\begin{subequations}
\begin{align}
[11 \vert 11] &= \frac{(2 - s^2)(11\vert11) - s^2(11\vert22)}{2(1 - s^2)} , \\
[11 \vert 22] &= \frac{(2 - s^2)(11\vert22)  - s^2(11\vert11)}{2(1 - s^2)} , \\
[11 \vert 12] &= 0, \\
[12 \vert 12] &= 0.
\end{align}
\end{subequations}
Hence, we find that the Mulliken approximation to the two-electron integrals for the H$_2$ molecule in a minimal basis, automatically leads to the extended Hubbard model used in our paper, with $U = [11 \vert 11]$ and $w = [11 \vert 22]$ in~\eqref{eq:LowdinTwoElectron}.

\section{Particle number consistency}
\label{ap:GWchemPot}
To have a consistent number of particles in the various approximations to the grand potential of the interacting system, we need to guarantee that we do not change the particle number when going from $\Omega_s$ to $\Omega$, so we require $N_s = N$. In other words, we need that
\begin{equation}
\frac{\du(\Omega - \Omega_s)}{\du\mu} = 0 .
\end{equation}
When inserting $\vec{G}_s$ in the Klein functional, this amounts to the requirement
\begin{multline}\label{eq:KleinConservN}
\trace\bigl\{(\vec{v}_{\text{ext}} - \vec{v}_s)\frac{\du\vec{\gamma}}{\du\mu}\bigr\} +
\frac{1}{\beta}\Trace\biggl\{\frac{\delta\Phi}{\delta\vec{G}_s}\frac{\du\vec{G}_s}{\du\mu}\biggr\} \\
= \frac{1}{\beta}\Trace\biggl\{\vec{\tilde{\Sigma}}\frac{\du\vec{G}_s}{\du\mu}\biggr\} = 0.
\end{multline}
Let us first consider the HF approximation. The HF self-energy is local in time, so $\vec{G}_s$ can be replaced by $\vec{\gamma}$. The derivative of the 1RDM with respect to the chemical potential is readily worked out as
\( \du\vec{\gamma}/\du\mu = \beta f_gf_u\vec{1} \),
so the particle conservation condition~\eqref{eq:KleinConservN} puts the following condition on the trace of the effective potential of the non-interacting reference system
\begin{equation}\label{eq:vsTrace}
\trace\bigl\{\vec{v}_s\bigr\}
= \trace\bigl\{\vec{\Sigma}^{\text{HF}} + \vec{v}_{\text{ext}}\bigr\}
= 4\biggl(\alpha + \frac{U + 2w}{2}\biggr) ,
\end{equation}
where we used the matrix elements of the HF self-energy~\eqref{eq:HFpotential} and have set $\tilde{n} = 1$. This exactly agrees with the chemical potential in the HF approximation~\eqref{eq:HFchemPot}, as this leads to $\epsilon_g^{\text{M}} + \epsilon_u^{\text{M}} = 0$.

Including the correlation part of the GW functional does not lead to a change in the trace of $\vec{v}_s$. This is most easily established by considering the derivative of $\Phi^{\text{GW}}_c$ in~\eqref{eq:PhiGWc} directly, as it yields
\begin{equation}
\frac{\du\Phi^{\text{GW}}_{\text{c},s}}{\du\mu}\biggr\rvert_{\crampedrlap{\tilde{n}=1}}
= -\frac{\beta u}{2}(2 - 2\tilde{n})\frac{\du\tilde{n}}{\du\mu}\biggr\rvert_{\crampedrlap{\tilde{n}=1}} = 0 .
\end{equation}
We find therefore, that the trace condition on $v_s$ remains the same as in~\eqref{eq:vsTrace} when using the GW approximation in the Klein functional.

In the Luttinger--Ward functional we have some additional terms~\eqref{eq:LWcor}. For the derivative of the linear term with respect to the chemical potential we find directly from~\eqref{eq:LWlinGW}
\begin{equation}
\frac{\du}{\du\mu}\Trace\bigl\{\vec{\tilde{\Sigma}}\vec{G}_s\bigr\}
\biggr\rvert_{\crampedrlap{\tilde{n}=1}}
= -\beta(2 - 2\tilde{n})u\frac{\du\tilde{n}}{\du\mu}\biggr\rvert_{\crampedrlap{\tilde{n}=1}} = 0.
\end{equation}
For the logarithmic term we do not have an explicit expression, but we can work it out as follows
\begin{multline}\label{eq:LWlnDmu}
-\frac{\du}{\du\mu}\Trace\bigl\{\ln\bigl(\vec{1} - \vec{G}_s\vec{\tilde{\Sigma}}\bigr)\bigr\} \\
= \Trace\biggl\{\bigl(\vec{1} - \vec{G}_s\vec{\tilde{\Sigma}}\bigr)^{-1}
\biggl(\frac{\du\vec{G}_s}{\du\mu}\vec{\tilde{\Sigma}} +
\vec{G}_s\frac{\du\vec{\Sigma}}{\du\mu}\biggr)\biggr\}
\end{multline}
The derivatives of the non-interacting Green's function with respect to the chemical potential is readily worked out in the frequency domain as
\begin{equation}
\frac{\du G_{s,kl}(\omega)}{\du\mu} = -\delta_{kl}\bigl(\omega - \epsilon^{\text{M}}_k\bigr)^{-2} ,
\end{equation}
which has the following useful property
\begin{equation}\label{eq:GsDmuSym}
\frac{\du G_{s,gg}(\omega)}{\du\mu} = \frac{\du G_{s,uu}(-\omega)}{\du\mu} .
\end{equation}
Likewise, the derivative of the GW self-energy with $\vec{G}_s$ inserted, can be calculated directly from its explicit expression in~\eqref{eq:SigmaGW}. For the gerade component we get
\begin{multline}
\frac{\du}{\du\mu}\Sigma^{\text{GW}}_{\text{c},s;gg/uu}[\vec{G}_s](\omega)
= \frac{\du}{\du\omega}\Sigma^{\text{GW}}_{\text{c},s;gg/uu}[\vec{G}_s](\omega) \\
{} - \frac{(u^2+v^2)(\omega_{u/g}^2 - \epsilon^2) -2\epsilon\tilde{f}(u^2 - v^2)}
{(\omega_{u/g}^2 - \zeta_+^2)(\omega_{u/g}^2 - \zeta_-^2)}\frac{\du\tilde{n}}{\du\mu} ,
\end{multline}
where we used $\omega_k = \omega - \epsilon^{\text{M}}_k$ as an abbreviation.
Due to the relation between the GW self-energy in~\eqref{eq:SigmaGWsym} for $\tilde{n} = 1$, the derivative with respect to the chemical potential obeys the following relation
\begin{align}\label{eq:SigmaDmuSym}
\frac{\du}{\du\mu}\Sigma^{\text{GW}}_{\text{c},s;gg}[\vec{G}_s](\omega)
&= \frac{\du}{\du\mu}\Sigma^{\text{GW}}_{\text{c},s;uu}[\vec{G}_s](-\omega) &
&\text{if $\tilde{n} = 1$.}
\end{align}
Note that~\eqref{eq:SigmaGWsym} also implies that if the trace of the single particle potential $\vec{v}_s$ satisfies~\eqref{eq:vsTrace}, that we also have
\begin{align}
\tilde{\Sigma}^{\text{GW}}_{\text{c},s;gg}[\vec{G}_s](\omega)
&= \tilde{\Sigma}^{\text{GW}}_{\text{c},s;uu}[\vec{G}_s](-\omega) &
&\text{if $\tilde{n} = 1$.}
\end{align}
Let us introduce the following abbreviation for the gerade component of the trace in~\eqref{eq:LWlnDmu} as
\begin{equation}
F(\omega) \isDefinedAs
\frac{\dfrac{\du G_{s,gg}(\omega)}{\du\mu}\tilde{\Sigma}_{gg}(\omega) +
G_{s,gg}(\omega)\dfrac{\du\tilde{\Sigma}_{gg}(\omega)}{\du\mu}}{1 - G_{s,gg}(\omega)\tilde{\Sigma}_{gg}(\omega)} .
\end{equation}
Using $G_{s,gg}(\omega) = -G_{s,uu}(-\omega)$ and the relations in~\eqref{eq:SigmaGWsym}, in~\eqref{eq:GsDmuSym} and in~\eqref{eq:SigmaDmuSym}, we find that the ungerade component is given by $-F(-\omega)$. The derivative of the logarithmic can therefore be written as
\begin{multline}
-\frac{\du}{\du\mu}\Trace\bigl\{\ln\bigl(\vec{1} - \vec{G}_s\vec{\tilde{\Sigma}}\bigr)\bigr\} \\
= \sum_p\bigl(F(\omega_p) - F(-\omega_p)\bigr) = 0 ,
\end{multline}
where we used that the direction of summation over the Matsubara frequencies $\omega_p$ is immaterial. Note that we could drop the convergence factor $\e^{\eta\omega_p}$ as the integrant already converges without it. So also for the Luttinger--Ward functional~\eqref{eq:vsTrace} is the appropriate trace of the single particle potential to keep $N = 2$.

\bibliographystyle{epj}
\bibliography{bible}

\end{document}